\documentclass[12pt,preprint]{aastex}

\slugcomment{Submitted to ApJ}

\shorttitle{The MUNICS Galaxy Luminosity Function}
\shortauthors{Drory et al.}

\newcommand{\M}{\ensuremath{\mathcal{M}}}%
\newcommand{\Lik}{\ensuremath{\mathcal{L}}}%
\newcommand{\Mmin}{\ensuremath{M_{\mathrm{min}}}}%
\newcommand{\Mmax}{\ensuremath{M_{\mathrm{max}}}}%
\newcommand{\zmin}{\ensuremath{z_{\mathrm{min}}}}%
\newcommand{\zmax}{\ensuremath{z_{\mathrm{max}}}}%
\newcommand{\Nobs}{\ensuremath{N_{\mathrm{obs}}}}%
\newcommand{\Nmodel}{\ensuremath{N_{\mathrm{model}}}}%
\newcommand{\Vmax}{\ensuremath{V_{\mathrm{max}}}}%
\newcommand{\Msun}{\ensuremath{\M_\odot}}%

\begin{document}


\title{The Munich Near-Infrared Cluster Survey (MUNICS) -- II. The
  $K$-band luminosity function of field galaxies to $z~\sim~1.2$}


\author{N.~Drory\altaffilmark{1}, 
R.~Bender\altaffilmark{2,3}, 
G.~Feulner\altaffilmark{2}, 
U.~Hopp\altaffilmark{2}, 
C.~Maraston\altaffilmark{3}, 
J.~Snigula\altaffilmark{2}} 

\and

\author{G.~J.~Hill\altaffilmark{1}}


\begin{abstract}
  We present a measurement of the evolution of the rest-frame $K$-band
  luminosity function to $z \sim 1.2$ using a sample of more than 5000
  $K$-selected galaxies drawn from the MUNICS dataset.  Distances and
  absolute $K$-band magnitudes are derived using photometric redshifts
  from spectral energy distribution fits to BVRIJK photometry.  These
  are calibrated using $> 500$ spectroscopic redshifts. We obtain
  redshift estimates having a rms scatter of $0.055$ and no mean bias.
  We use Monte-Carlo simulations to investigate the influence of the
  errors in distance associated with photometric redshifts on our
  ability to reconstruct the shape of the luminosity function.
  Finally, we construct the rest-frame $K$-band LF in four redshift
  bins spanning $0.4<z<1.2$ and compare our results to the local
  luminosity function. We discuss and apply two different estimators
  to derive likely values for the evolution of the number density,
  $\Phi^*$, and characteristic luminosity, $M^*$, with redshift.
  While the first estimator relies on the value of the luminosity
  function binned in magnitude and redshift, the second estimator uses
  the individually measured $\{M,z\}$ pairs alone. In both cases we
  obtain a mild decrease in number density by $\sim 25\%$ to $z=1$
  accompanied by brightening of the galaxy population by $0.5$ to
  $0.7$~mag. These results are fully consistent with an analogous
  analysis using only the spectroscopic MUNICS sample.  The total
  $K$-band luminosity density is found to scale as $d\log\rho_L / dz =
  0.24$.  We discuss possible sources of systematic errors and their
  influence on our parameter estimates. By comparing the luminosity
  density and the cumulative redshift distributions of galaxies in
  single survey fields to the sample averages, we show that cosmic
  variance is likely to significantly influence infrared selected
  samples on scales of $\sim 100$ square arc minutes.
\end{abstract}


\altaffiltext{1}{University of Texas at Austin, Austin,
  Texas 78712, \{drory,hill\}@astro.as.utexas.edu}

\altaffiltext{2}{Universit\"ats-Sternwarte M\"unchen, Scheinerstra\ss
  e 1, D-81679 M\"unchen, Germany,
  \{bender,feulner,hopp,snigula\}@usm.uni-muenchen.de}

\altaffiltext{3}{Max-Planck Institut f\"ur
  extraterrestrische Physik, Giessenbachstra\ss e, Garching, Germany,
  \{bender,maraston\}@mpe.mpg.de}


\section{Introduction}
\label{sec:introduction}

The luminosity function (LF) is the most basic statistic used to study
galaxy populations and their evolution. Its dependence on wavelength
and look-back time provides important constraints on the evolution of
the global properties of the galaxy population.

Recently, progress in measuring the optical galaxy LF was made both
locally and at higher redshift. Locally, results from the 2dF survey
\citep{2dF99,2dF02} and 2MASS \citep{Kochaneketal01} provide much
improved measurements of the LF as a function of galaxy morphology,
environment, and wavelength. In spite of some controversy regarding
the normalization and the very faint end slope of the local LF, it is
a well established result that the LF depends on galaxy type, although
the correlations are not always very tight. Generally, the faint end
is populated by galaxies of smaller mass, later morphology, bluer
colors and later spectral type, and stronger line emission.  The
bright end is dominated by early-type spirals and ellipticals, and the
very bright end by giant ellipticals
\citep[e.g.][]{CFA94,BPLK98,2dF02}.

A number of studies of the luminosity function of galaxies at $0 \la z
\la 1$ using optically selected redshift surveys have been published
in recent years
\citep[e.g.][]{EEP88,CFRS695,HCEB97,LYCE97,LGHO98,UKST98,CNOC299}.
These surveys (with samples of typically hundreds of objects) have
consistently found similar trends in the evolution of the rest-frame
$B$-band field galaxy LF. The main result was the contrast between the
rapid evolution of the blue, star-forming sub-population and the mild
change in the redder, early-type population. Simultaneously, first
results at $z \sim 3$ became available through the study of
Lyman-break galaxies \citep{SSADGP01}.

The ground-breaking CFRS \citep{CFRS95} found that the luminosity
function in the rest-frame $B$-band of the blue field population
brightens by roughly 1~mag to $z \simeq 1$ and also becomes steeper at
the faint end at $z \ga 0.5$ \citep[see also][]{HCEB97}. In contrast,
the red population was found to show very little change in either
number density or luminosity \citep{CFRS695}.  \citet{CNOC299}
attempted to discriminate explicitly between number-density and
luminosity evolution in the CNOC2 sample consisting of roughly 2000
galaxies.  They found that the early-type population shows positive
luminosity evolution (1.6 mag) which is nearly compensated by negative
density evolution (factor of 0.5), so that there is little net change
in their overall LF. The intermediate-type population shows positive
luminosity evolution (0.9 mag) plus weak positive density evolution
(factor of 1.7), resulting in mild positive evolution in its
luminosity density.  The amount of luminosity evolution in the
early-type and intermediate-type populations was found to be
consistent with expectations from models of passive evolution of their
stellar populations. In contrast, the late-type population is best fit
by a strong increase in number density at high redshift (factor of
4.1), accompanied by only little positive luminosity evolution (0.2
mag).  The overall $B$-band luminosity density of late-type objects
was found to increase rapidly in both the CFRS and CNOC2 samples,
while the luminosity density of early-type objects was found to be
nearly constant.

In contrast to the blue and optical wave-bands, the near infrared
$K$-band light of galaxies is much less affected by dust, much less
sensitive to ongoing star-formation and much less dependent on galaxy
type. Furthermore, the near-IR luminosity is as a reasonable tracer of
the stellar mass of a galaxy, since the near-IR light is dominated by
the light of old and evolved stars.  Therefore, by investigating the
$K$-band properties of galaxies as a function of redshift, we may hope
to be able to move from a picture dominated by the evolution of star
formation to one which focuses on the assembly history of mass in
these systems -- one of the most fundamental predictions of cold dark
matter based structure formation theories.

Most near-IR selected surveys so far have been either small in size or
shallow \citep[e.g.][]{GPMC95,GSFC97,CSHC96,SSCM98}.  Furthermore,
optical followup spectroscopy of near-IR selected samples to $z \sim
1$ is difficult because of the large range of optical to near-IR
(e.g.\ $R\!-\!K$) colors of galaxies, although spectroscopic samples
of a few hundred objects are now available
\citep{MUNICS5,K20-03,K20-02}. If any change in the LF was seen at
all, only mild luminosity evolution at $z>0.5$ has been significantly
detected.

In this work, we aim at measuring the rest-frame $K$-band luminosity
function of galaxies and its evolution with redshift in the range $0.4
< z < 1.2$ using the $K$-band selected, BVRIJK multicolor data set of
the MUNICS project \citep{MUNICS1}. The sample contains more than 5000
objects and covers a large solid angle. We use spectroscopically
calibrated photometric redshifts (using more than 500 spectroscopic
redshifts) to estimate distances and spectral energy distribution
(SED) fitting techniques to extrapolate the SED to rest-frame $K$.

This paper is organized as follows. First, we briefly introduce the
galaxy sample we use in this work in Sect.~\ref{sec:galaxy-sample}.
Next, in Sect.~\ref{sec:phot-redsh}, we describe the photometric
redshift estimation technique along with the construction of a set of
semi-empirical SEDs best matching the SEDs present in the sample.
In Sect.~\ref{sec:comp-lumin-funct} we discuss the construction of the
rest-frame $K$-band luminosity function from photometric redshifts and
assess the influence of errors in $z$ by means of Monte-Carlo
simulations. The resulting luminosity function is presented in
Sect.~\ref{sec:lf-results}, and its evolution is investigated in
Sect.~\ref{sec:likel-analys-lumin}.  Sect.~\ref{sec:discussion}
discusses the results and Sect.~\ref{sec:summary} summarizes this
work.

We assume $\Omega_{\rm M} = 0.3$, $\Omega_{\Lambda} = 0.7$ throughout
this paper. We write Hubble's Constant as $H_0 = 100\ h\ \mathrm{km\ 
  s^{-1}\ Mpc^{-1}}$, using $h = 0.65$ unless the quantities in
question can be written in a form explicitly depending on $h$.


\section{The galaxy sample}
\label{sec:galaxy-sample}

MUNICS is a wide-area, medium-deep, photometric and spectroscopic
survey selected in the $K$ band, reaching $K \sim 19.5$. It covers an
area of roughly one square degree in the $K$ and $J$ bands with
optical follow-up imaging in the $I$, $R$, $V$, and $B$ bands in 0.4
square degrees. \citet[][hereafter MUNICS~I]{MUNICS1} discusses the
field selection, object extraction and photometry in the $K,J,I,R$,
and $V$ bands. The $B$ band imaging was completed more recently and
the data were processed in an analogous way as described in MUNICS~I.
Detection biases, completeness, and photometric biases of the MUNICS
data are analyzed in detail in \citet[][hereafter MUNICS~IV]{MUNICS4}.

The MUNICS photometric survey is complemented by spectroscopic
follow-up observations of all galaxies down to $K\le 17.5$ in 0.25
square degrees, and a sparsely selected deeper sample down to $K \le
19$.  It contains more than 550 secured redshifts. The spectra cover a
wide wavelength range of $4000-8500$\AA\ at $13.2$\AA\ resolution, and
sample galaxies at $0<z<1$. These observations are described in detail
in \citet[][hereafter MUNICS~V]{MUNICS5}.

The galaxy sample used in this work is a subsample of the MUNICS
survey Mosaic Fields (see MUNICS~I), selected for best photometric
homogeneity, good seeing, and similar depth. Furthermore, in each of
the survey patches, areas close to the image borders in any passband,
areas around bright stars, and regions suffering from blooming are
excluded. The subsample covers 0.28 square degrees in $B$, $V$, $R$,
$I$, $J$, and $K$.  Table~\ref{t:good-fields} lists the Mosaic Fields
used in the subsequent analysis. It is almost identical to the sample
used in \citet[][hereafter MUNICS~III]{MUNICS3}, except for the
additional $B$-band imaging data (and hence refined photometric
redshifts).

Stars are identified following the procedure described in MUNICS~I,
adding a color criterion in the $J\!-\!K$ vs. $V\!-\!I$ plane to
exclude faint stellar sources which cannot be separated from galaxies
morphologically (see Fig.~10 in MUNICS~I). This color criterion may
also exclude $z \la 0.2$ compact blue galaxies.  Such galaxies are
very unlikely to be present in the $K$-selected sample, given our
magnitude limit.  We will restrict our analysis of the luminosity
function to $z > 0.4$ such that this is anyway not a problem.

To make sure that our star-galaxy separation does not cause any
systematic biases in our analysis, we double check the method by
including stellar SEDs in the template library used to estimate
photometric redshifts (see below). We find that the objects identified
as stars are indeed better fit by a pure stellar SED than by any
galactic SED (see right panel in Fig.~\ref{f:phot-z-example}).  Most
exceptions (a few percent of the stars) are most probably binaries.
Using only the SED fit to discriminate stars from galaxies does not
change any of the results of this work.


\section{Photometric redshifts}
\label{sec:phot-redsh}

Photometric redshifts are derived using the method presented in
\citet{photred00} and \citet{Bender03}. We only briefly outline the
procedure in this section. The method is a template matching algorithm
rooted in Bayesian statistics closely resembling the method presented
by \citet{Benitez00}.  However, instead of relying on a predetermined
set of template SEDs, semi-empirical templates matching the
photometric properties of the sample are used. The templates are
derived by fitting stellar population models of \citet{Maraston98} of
different age and dust extinction and Kinney-Calzetti
\citep{KinneyCalzetti} spectra to combined broad-band energy
distributions of MUNICS and FDF \citep{FDF1} galaxies having
spectroscopic redshifts.  In this way, representative galaxy templates
of mixed stellar populations (variable age, metallicity, and dust
extinction) optimized for the MUNICS dataset are obtained.

The total spectroscopic sample is divided into 2 groups of objects.
The first subsample (all objects with redshifts in the field S2F1) is
used for constructing SED templates, the second subsample (all other
objects with spectroscopic redshifts) is used for comparing
spectroscopic and photometric redshifts and thereby calibrating the
SED library.

\begin{figure}
  \plottwo{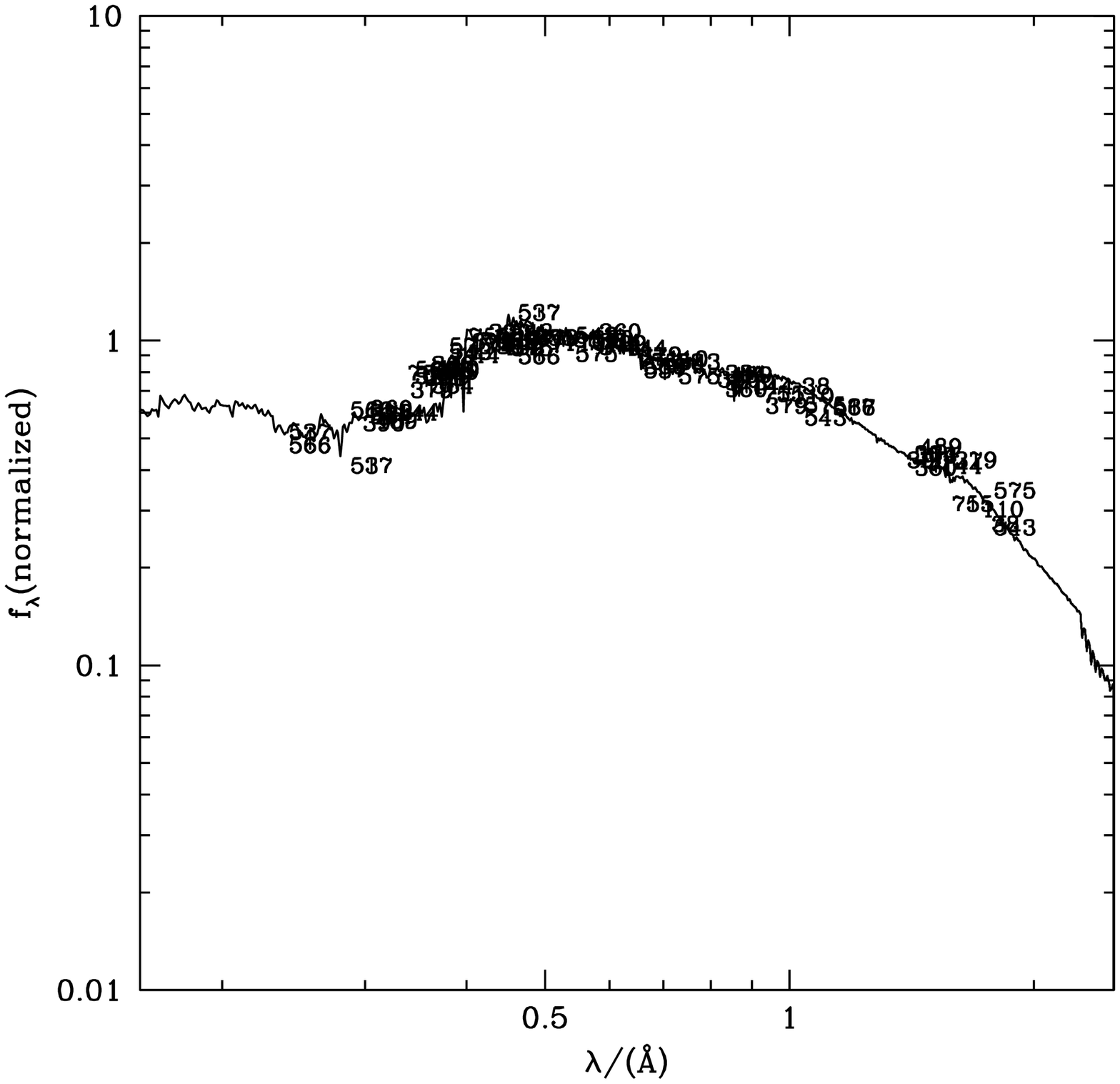}{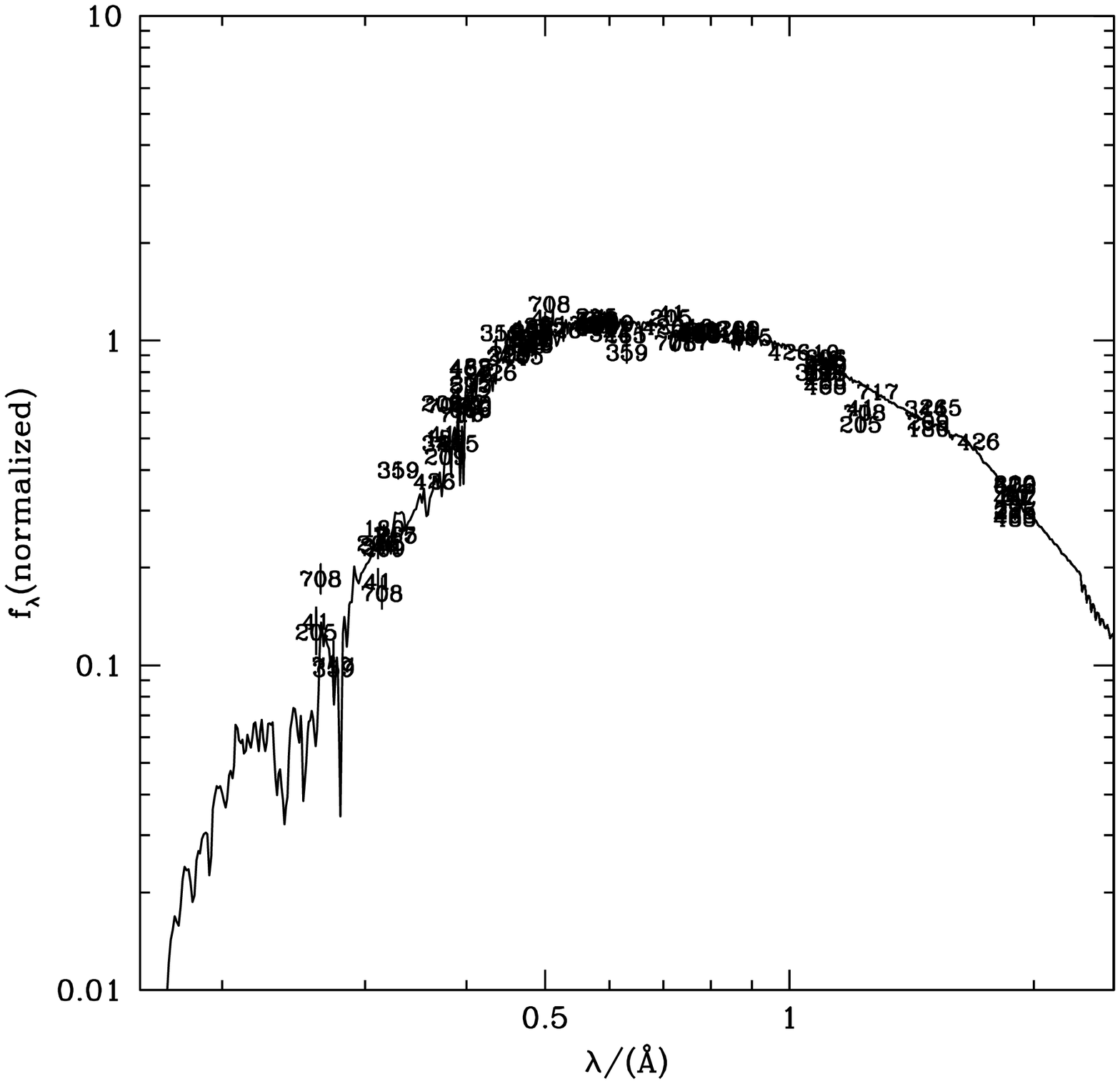}
  \caption{A low-resolution spectrum constructed
    from BVRIJK photometry of objects with known spectroscopic
    redshifts. The objects' photometry is de-redshifted to $z=0$ and
    normalized to a common magnitude and fit by stellar population
    synthesis models. Objects best fitting the same model are grouped
    together and used in an iterative procedure to find an optimal set
    of SEDs to use as templates. The measured fluxes of each object
    are marked by the object's ID. The solid line is the final SED
    template which best fits this class of objects.}
  \label{f:sed-fit}
\end{figure}

The observed-frame apparent magnitudes of objects in the first
subsample are transformed to rest-frame redshift zero, and fitted by
an initial set of stellar population synthesis models. Objects best
fitting the same model are grouped together. Since each group will
contain objects from a variety of redshifts, a densly sampled SED from
the broad band photometry of these objects is obtained. This procedure
is illustrated in Fig.~\ref{f:sed-fit}. 

This initial set of SEDs is used to determine photometric redshifts
for the total sample of objects having spectroscopic redshifts. The
photometric redshifts are compared to the spectroscopic ones, and,
additionally, the same de-redshifting procedure is applied to the
spectroscopic sub-sample not used for the initial construction of the
SEDs, only that now we group the objects by the SED that gave the best
fit during the determination of the photometric redshift. Using this
comparison, deficiencies in the set of SEDs can be identified as those
become apparent through systematic offsets between the de-redshifted
magnitudes and the SED templates. This is the case since such
deficiencies lead to a wrongly determined photometric redshift and
therefore to the assignment of a wrong SED.

\begin{figure}
  \epsscale{0.6}
  \plotone{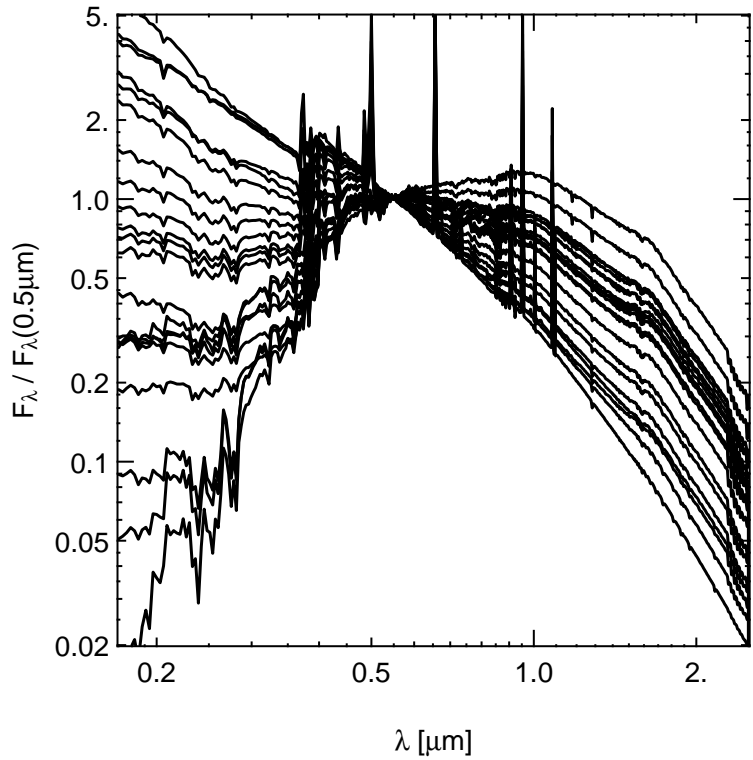}
  \caption{Final SED library used to determine photometric redshifts. 
    Note that there are no strongly star forming (young) SEDs in the
    final set, as can be expected in a $K$-band selected sample. Also
    noteworthy is the fact that we definitely need dust-reddened
    SEDs.}
  \label{f:model-seds}
\end{figure}

This procedure is repeated with a refined set of SEDs, by changing
SEDs, abolishing some and adding others, until a satisfactory library
of template SEDs is found.  Fig.~\ref{f:model-seds} shows the final
template SED library used to derive photometric redshifts in what
follows.

\begin{figure}
  \epsscale{0.6}
  \plotone{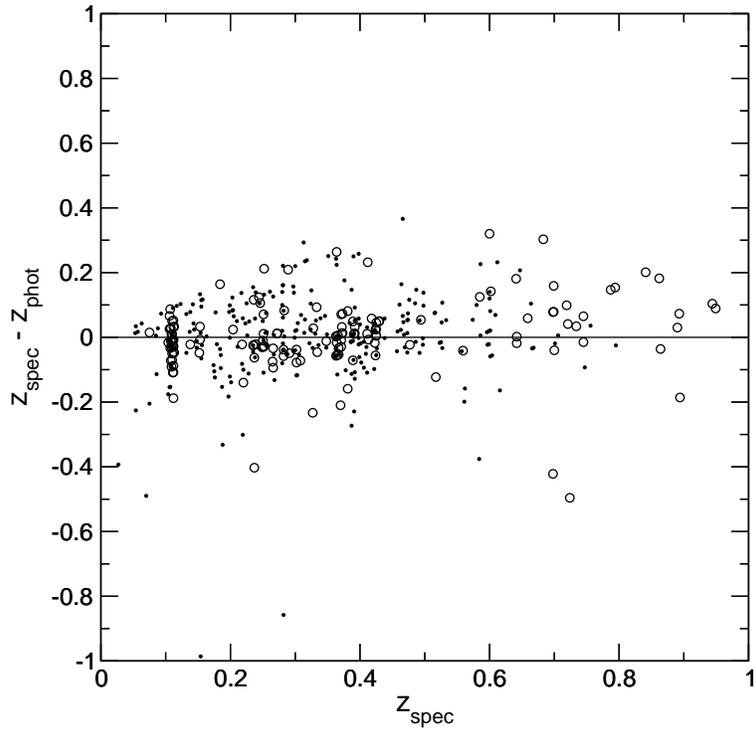}
  \caption{The difference between spectroscopic and photometric redshift 
    vs.\ spectroscopic redshift for the subsample used to construct
    the SED templates (open circles) and all other objects (filled
    circles).}
  \label{f:zzcomp}
\end{figure}

In Fig.~\ref{f:zzcomp} we plot the difference between spectroscopic
and photometric redshift vs.\ spectroscopic redshift for the subsample
used to construct the SED templates and for the rest of the sample
which was used to test the procedure. There is no apparent difference
between these two error distributions.

\begin{figure}
  \epsscale{1}
  \plottwo{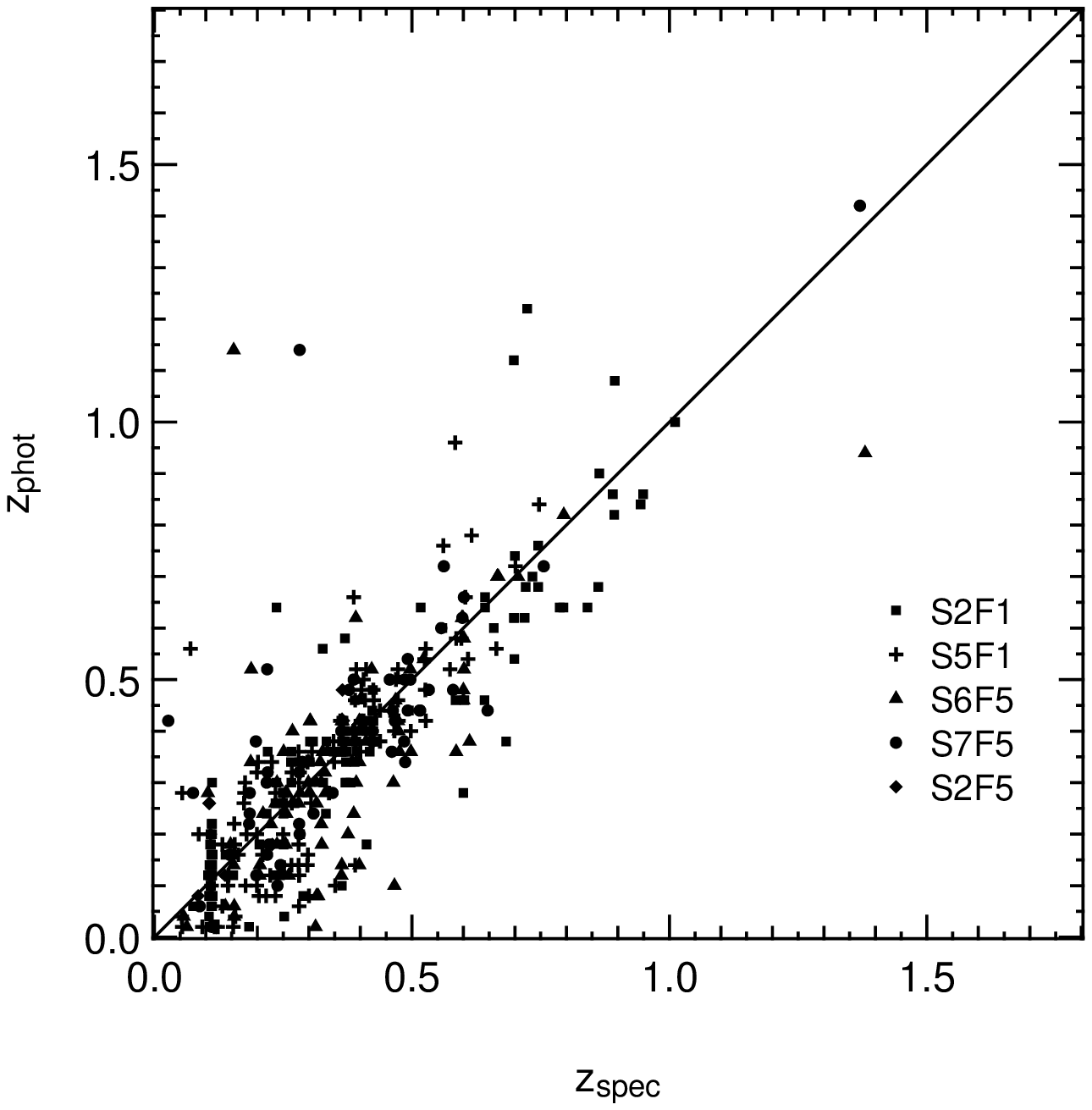}{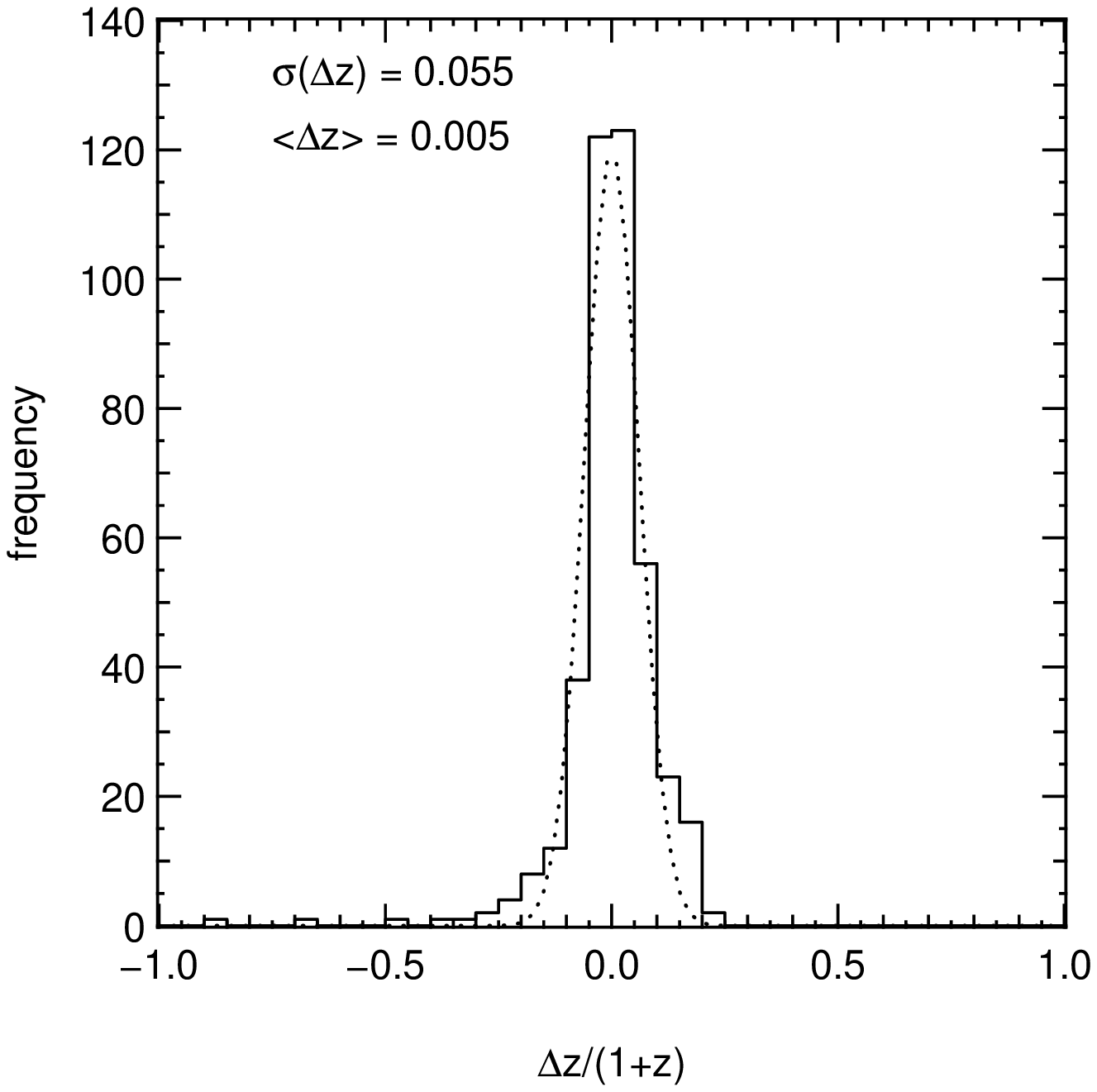}
  \caption{Left panel: Comparison of photometric and spectroscopic redshifts
    for $\sim 500$ objects in five survey patches (different symbols).
    Right panel: The histogram of the redshift errors. The rms scatter
    is approximately Gaussian (dotted line: best-fit Gaussian) of a
    width $\sigma$~=~0.055 and an insignificant mean deviation from
    the unity relation of $\langle\Delta z\rangle = -0.005$.}
  \label{f:zz}
\end{figure}

Finally, Fig.~\ref{f:zz} compares photometric and spectroscopic
redshifts for all $\sim 500$ objects within five MUNICS Mosaic Fields
and shows the distribution of redshift errors.  The typical scatter in
the relative redshift error $\Delta z / (1+z)$ is 0.055. The mean
redshift bias is negligible.  The distribution of the errors is
roughly Gaussian. There is no visible difference between the
distributions among the survey fields. Although this performance is
encouraging, it is important to say that the spectroscopic data become
sparse at $z \ga 0.6$ and there are only two spectroscopic redshifts
at $z > 1$.

\begin{figure}
  \epsscale{0.32}
  \plotone{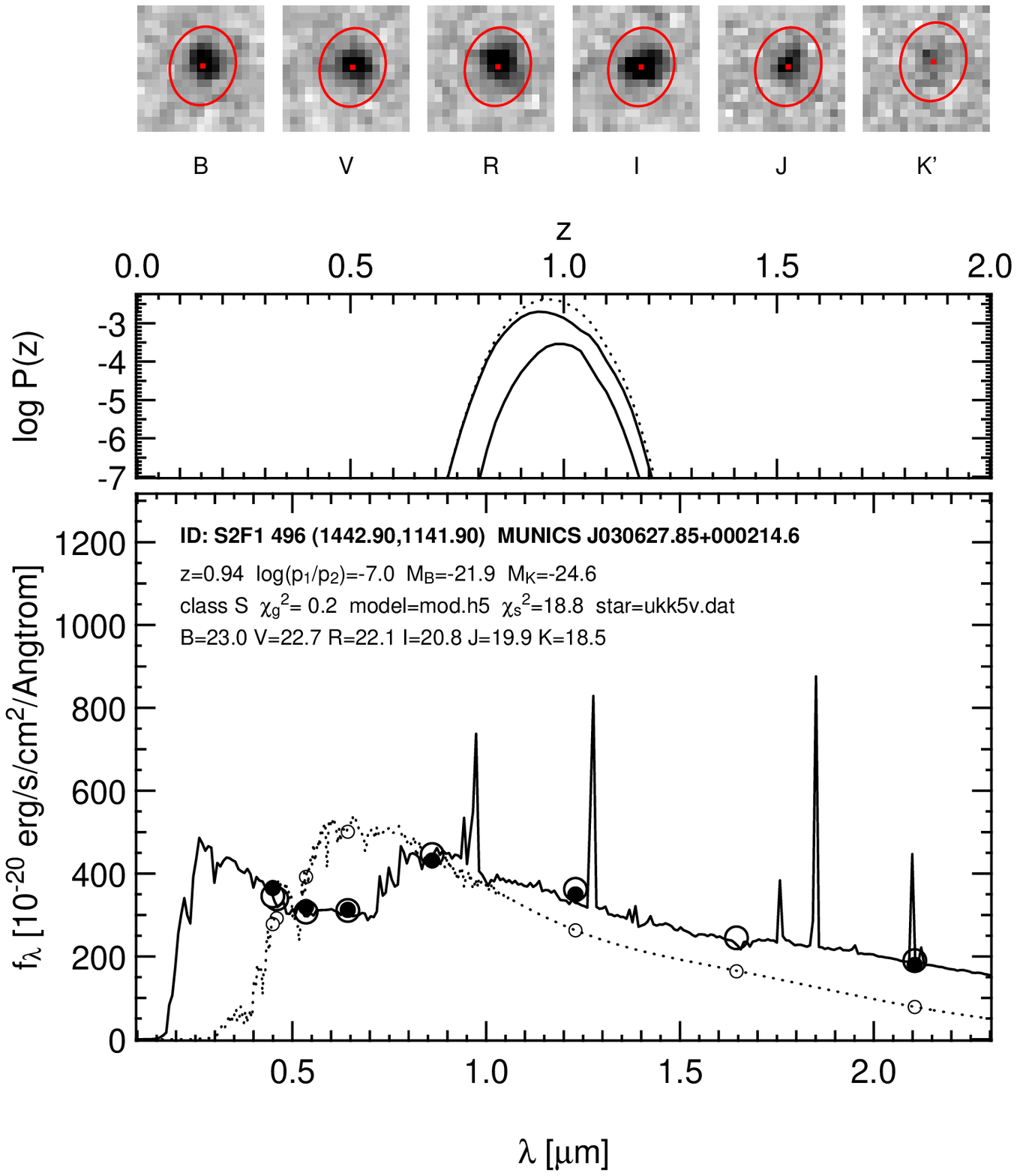}
  \plotone{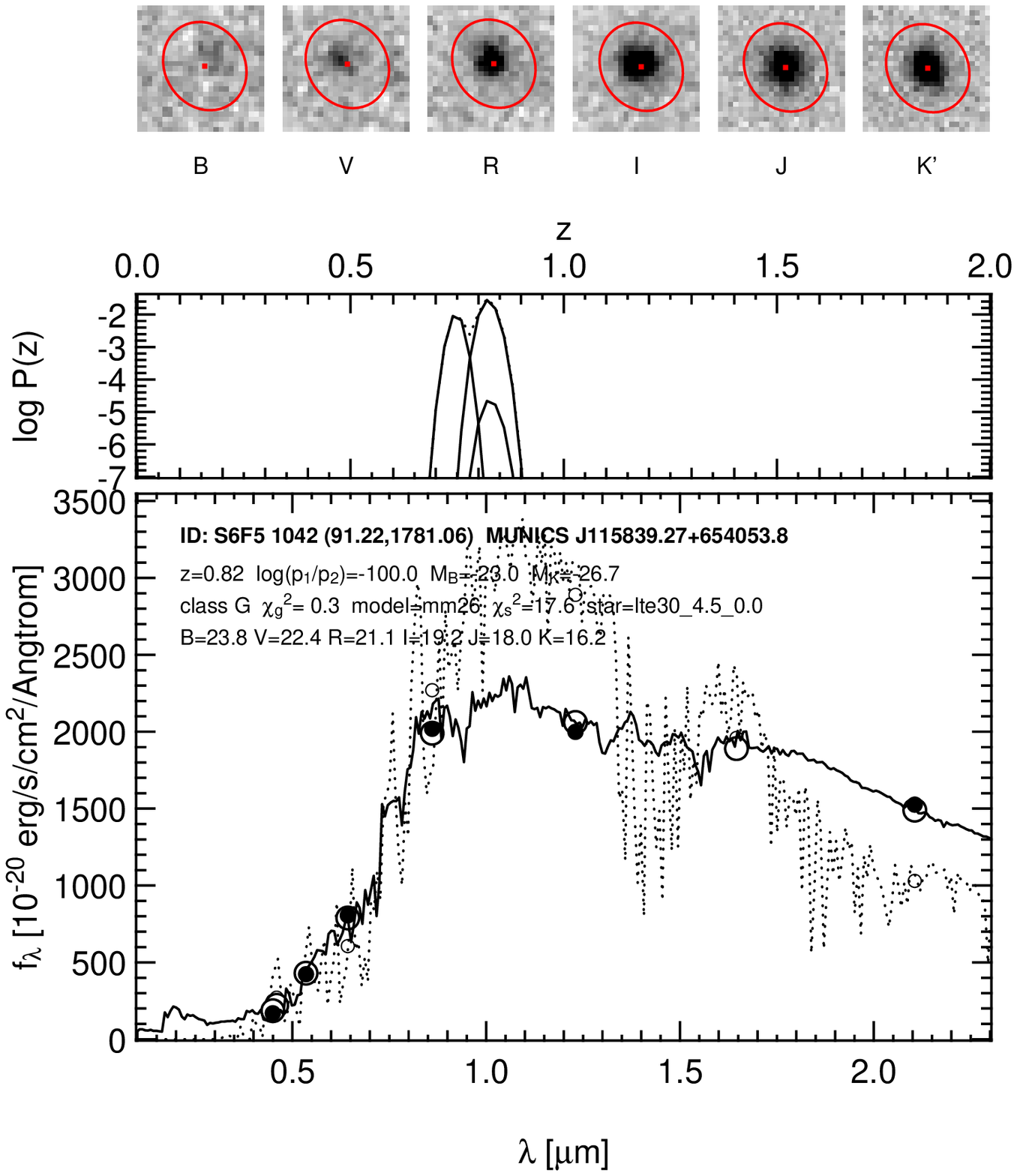}
  \plotone{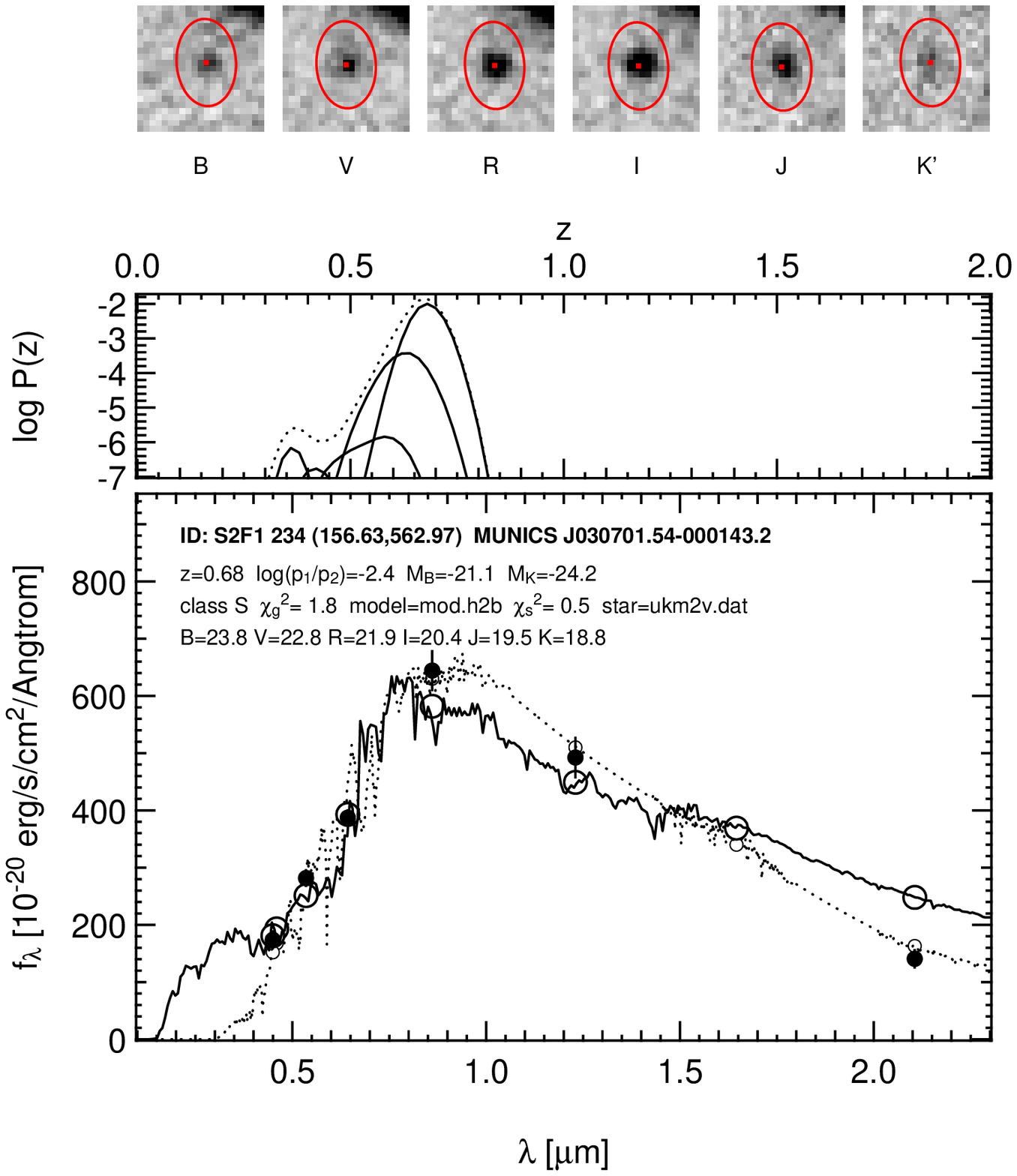}
  \caption{Examples of the determination of photometric redshifts. 
    The implications of these examples are discussed in the text. The
    lower panels show the best-fitting galaxy SED (solid line), the
    best-fitting stellar SED (dotted line), the expected fluxes in
    observed-frame $B$,$V$,$R$,$I$,$J$, and $K$ (open circles) and the
    measured fluxes (filled circles). The middle panel shows the
    redshift probability function for the different galaxy template
    SEDs and the total redshift probability function. The upper panels
    show thumbnail images of the objects in our six pass bands.}
  \label{f:phot-z-example}
\end{figure}

To illustrate the use of photometric redshifts, we show two
instructive examples of photometric redshift determinations of
galaxies and the identification of an M-star in
Fig.~\ref{f:phot-z-example}. These examples help to understand how the
technique works and the uncertainties involved.

Firstly, a spiral-like system at redshift around unity. Here the
Balmer break is redshifted beyond the $R$-band filter, and only one
SED contributes significantly to the global peak in the redshift
probability distribution. The rather broad probability distribution in
redshift is due to the fact that the Balmer break is in the rather
large sampling gap between the $R$ and $I$ bands, and therefore its
position is not very well determined.  Note that this object, although
very bright in the optical, is very faint in the $K$ band, and
therefore probably a rather low-mass system.

Secondly, an early-type object at redshift $z \sim 0.82$. The redshift
determination can be regarded as quite secure although there are
competing SEDs around the global peak of the redshift probability
function. In this case, the redshift is determined by the 4000\AA\ 
break and the steep decline in flux bluer thereof.  Because the object
is undetected in $B$ and only barely detected in $V$, the rest-frame
UV and blue slopes of the spectrum are not firmly determined and hence
slightly differing effective ages are giving reasonable fits. This
uncertainty in age or type translates into an uncertainty in redshift.
Therefore there are competing SEDs at similar but not identical
redshifts contributing to the total redshift probability distribution.

Common to both above examples is that the position in redshift of the
global maximum of the total redshift probability distribution is
always compatible with the redshift one would derive by looking at the
probability distribution of the most likely SED.

The third panel in Fig.~\ref{f:phot-z-example} shows an object best
fit by the SED of an M2 star. We have used a criterion based on a
comparison of the best $\chi^2$ for redshifted galaxy template SEDs
and stellar SEDs to discriminate between star and galaxies, to test
the robustness of the morphological and color based star-galaxy
separation presented in MUNICS~I, hoping to be able to improve the
procedure at faint magnitudes. The main problem hereby is the lack of
reliable SEDs for cool stars with the necessary broad wavelength
coverage. Using only this SED based star-galaxy separation instead of
the morphological method does not change the results of this work.
Note that we present an independent measure of the reliability of our
star-galaxy separation in MUNICS~V, where we test it against blind
spectroscopy.

\begin{figure}
  \epsscale{0.6}
  \plotone{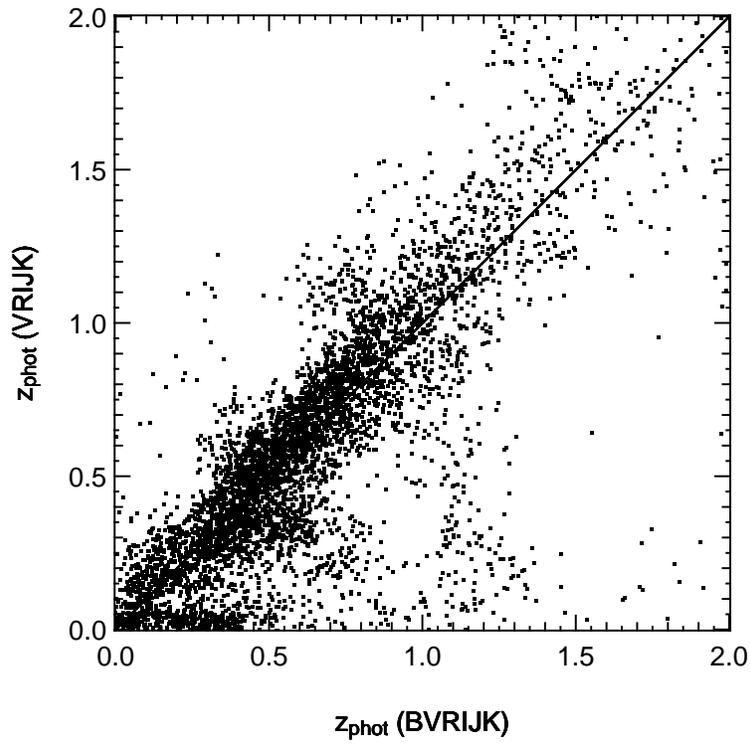}
  \caption{Comparison of photometric redshifts from the older VRIJK MUNICS
    sample used in MUNICS~III with the current BVRIJK sample used in
    this work.}
  \label{f:zold-znew}
\end{figure}

It is interesting to compare the photometric redshifts obtained using
the present six-color photometry to the ones used in MUNICS~III, where
we did not have B-band imaging yet and the spectroscopic calibration
sample was smaller (310 objects as opposed to 550 now). This
comparison is shown in Fig.\ref{f:zold-znew}. In spite of overall
reasonable agreement, there are some important differences. The most
notable one is that objects tend to scatter to lower redshifts if
$B$-band photometry is missing. Objects are scattered preferably to
the region in redshift below which the 4000\AA\ break enters the
observed filter set (because this peak in the $z$-probability function
cannot be excluded by the missing blue photometry). Another set of
objects is redistributed from $z \sim 1.2$ (six-colors) to lower
redshifts (five-colors). These objects all have blue colors and rather
late type SEDs. Both effects are small in numbers, affecting $\sim 3$
per cent of the sample. Therefore they do not show up in the
spectroscopic calibration and are difficult to eliminate
systematically. The influence of such an effect on the luminosity
function can be significant although the number of affected objects is
small if rare high-luminosity objects are preferably affected. This is
discussed below in Sect.~\ref{sec:lf-results}.

\begin{figure}
  \epsscale{1}
  \plottwo{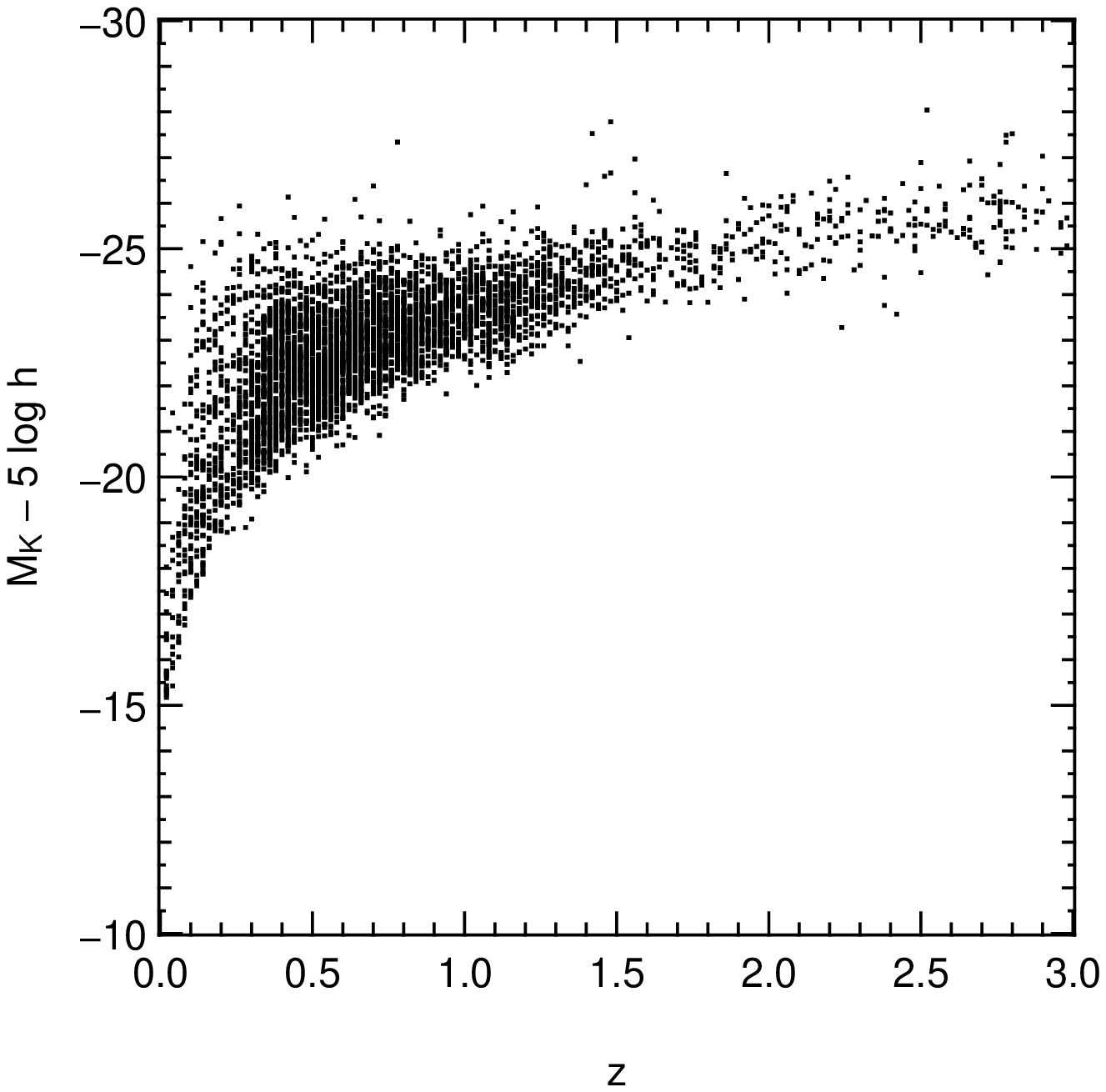}{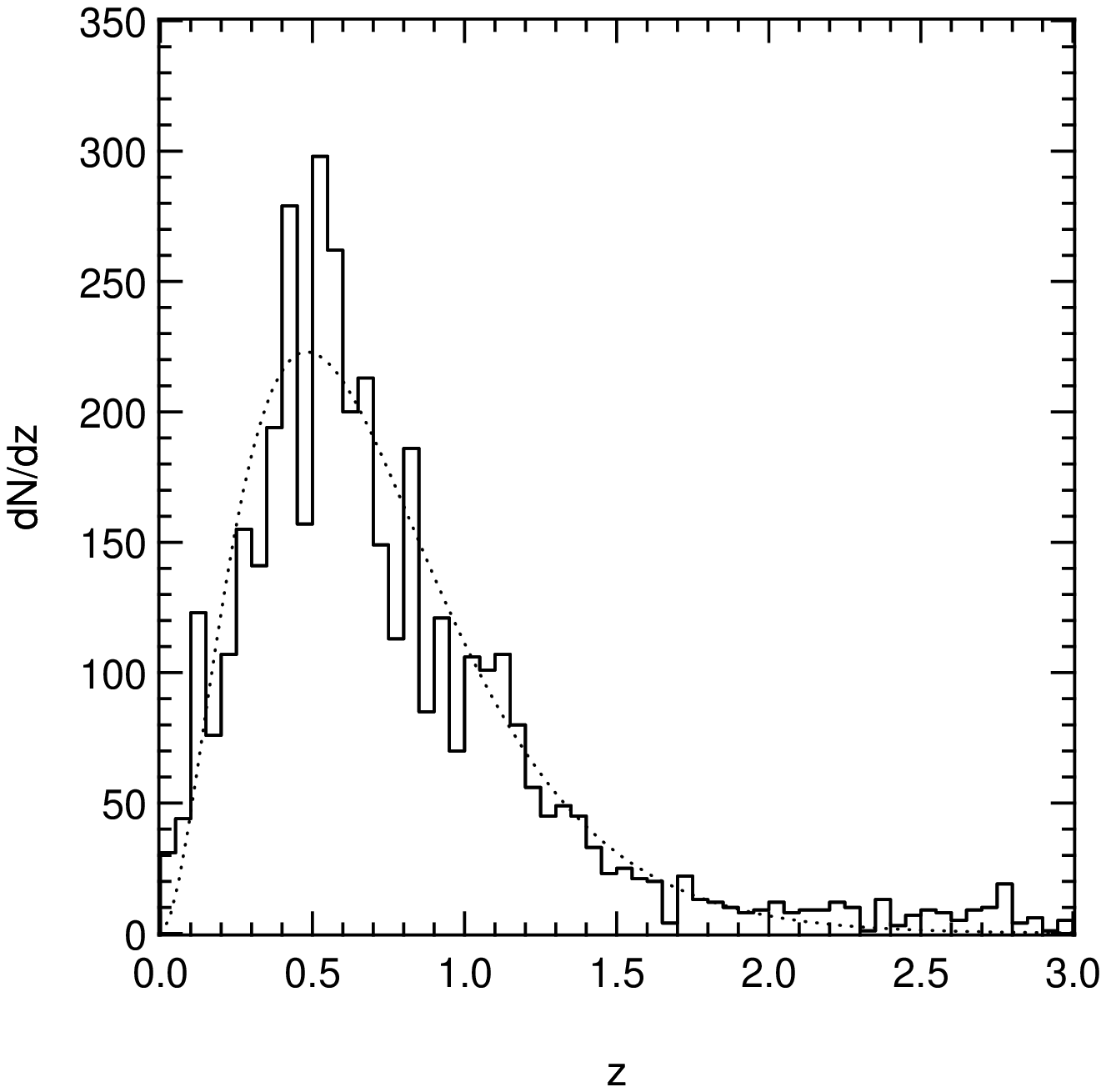}
  \caption{Left panel: the $M_K$ -- $z$ relationship for the total 
    MUNICS sample of 5132 galaxies using photometric redshifts. Right
    panel: the photometric redshift distribution of the sample and a
    best-fit analytic description (dotted line; see text). The
    distribution of spectroscopic redshifts is also shown for
    comparison (dashed line).}
  \label{f:mkz}
\end{figure}

In Fig.~\ref{f:mkz} we plot absolute $K$-band magnitude vs.\ 
photometric redshift and the redshift distribution of the total sample
discussed here, containing 5132 galaxies. The redshift distribution
peaks around $z \approx 0.5$ and has a tail extending to $z \approx
3$. An analytical fit of the form

\begin{equation}
  \label{eq:dndz-analytic}
  \frac{dN}{dz} \; = \; \frac{\beta z^2}{\Gamma(3/\beta)\: z_0^3}\: 
  e^{-(z/z_0)^{\beta}}
\end{equation}
is also shown. The best-fitting values are $z_0 = 0.101, \beta =
0.746$.

\begin{figure}
  \epsscale{0.6}
  \plotone{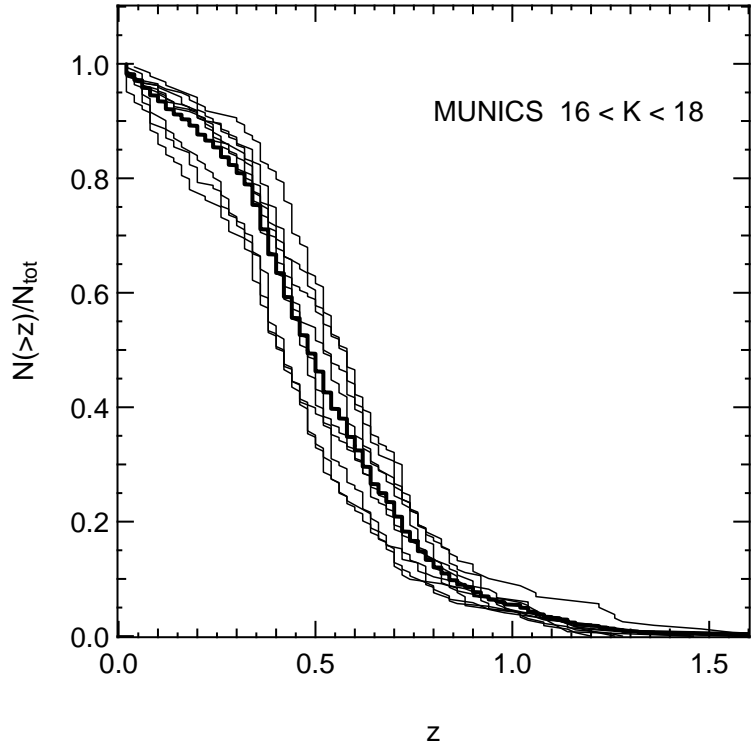}
  \caption{Cumulative redshift distribution of MUNICS galaxies with 
    apparent $K$-band magnitudes $16 < m_K < 18$. The thin lines are
    data from individual survey fields, the thick line is the total
    sample.}
  \label{f:cumul_z}
\end{figure}

Finally, we plot the cumulative redshift distribution of bright
galaxies in the $K$-band in Fig.~\ref{f:cumul_z}. We select galaxies
with apparent $K$-band magnitudes in the interval $16 < m_K < 18$,
where the present sample is definitively complete. The figure also
compares the data in the single survey fields to the sample average.
We find that cosmic variance is significant on scales of $\sim 100$
square arc minutes, the typical size of previous infrared selected
surveys.


\section{Computing the luminosity function}
\label{sec:comp-lumin-funct}

\subsection{The \Vmax\ method}
\label{sec:vmax-method}

We estimate the luminosity function, $\Phi(M)dM$, the comoving number
density of galaxies with absolute magnitude in the range $[M, M+dM)$,
using the \Vmax\ formalism \citep{Schmi68} to account for the fact
that some fainter galaxies are not visible in the whole survey volume.

Each galaxy in a given redshift bin $[z_l,z_h)$ contributes to the
number density an amount inversely proportional to the volume in which
the galaxy is detectable in the redshift bin given all relevant
observational constraints:
\begin{equation}
  V_i \; = \; d\Omega \,
  \int_{z_l}^{\mathrm{min}(z_h,z_{max})}\,\frac{dV}{dz}\,dz,
\end{equation}
where $dV/dz$ is the comoving volume element, $d\Omega$ is the survey
area, $z_{max}$ is the maximum redshift at which galaxy $i$ having
absolute magnitude $M_{K,i}$ is still detectable given the limiting
apparent magnitude of the survey and the galaxy's SED (the best-fit
SED from the photometric redshift determination in our case).

Additionally, the contribution of each galaxy $i$ is weighted by the
inverse of the detection probability, $P(m_{K,i})$, where we assume
that the detection probability is independent of the galaxy type and
can be approximated by that of point-like sources. We only include
objects with $P(m_{K,i}) > 0.75$, such that this correction is always
small.  We have checked that this correction does not bias our results
by comparing to what we get for higher completeness limits. Also, the
results of completeness simulations discussed in MUNICS~IV demonstrate
that although there exist profile-dependent surface-brightness
selection biases, these are under control for redshifts up to $z \ga
1$, and that the onset of incompleteness is roughly independent of
profile type.

The comoving number density of objects in a given absolute magnitude
interval and redshift bin is finally calculated as
\begin{equation}
  \Phi(M)dM = \sum_i \frac{1}{\Vmax^i}\frac{1}{P(m_{K,i})} dM,
\end{equation}
where the sum is to be taken over all objects $i$ in the bin.

The advantage of the \Vmax\ method is that it is non-parametric, i.e.\ 
no assumption on the form of the LF is made and the estimate of the LF
in each redshift and magnitude bin is independent. There is no need to
compute the normalization of the LF separately. It is quick and easy
to compute.  The major disadvantage is that this method is sensitive
to clustering in the sample. In our case we might hope that this is
not a major concern since we average over a large volume and probe
multiple independent lines of sight.

\citet{TYI00} performed a systematic comparison of different
estimators for the luminosity function, finding that the \Vmax\ 
estimator yields a completely unbiased result if there is no
inhomogeneity. Earlier claims that the \Vmax\ estimator is biased even
without clustering were not confirmed. The \Vmax\ estimator was also
found to give consistent results with other statistical estimators
analyzed, despite of its sensitivity on large scale structure.

The last missing ingredient is the absolute $K$-band magnitude. It is
computed by extrapolation from the observed-frame colors, using the
best-fit SED from the photometric redshift code. The near-IR slopes of
the SEDs are fairly uniform across galaxy types (the $K$-band
$k$-corrections are small and almost type-independent) and so the
uncertainty introduced by this extrapolation is rather small, of the
order of $\Delta M_K \sim 0.1$ mag in the mean, and thus small
compared with the uncertainty of the total rest-frame $K$-band
magnitude coming from the uncertainty in the distance.

\subsection{Monte-Carlo simulations}
\label{sec:monte-carlo-simul}

The use of photometric redshifts, in general, can introduce systematic
errors in the derived galaxy distances and therefore in the luminosity
function which we want to derive in what follows. Therefore, we
perform Monte-Carlo simulations to test the robustness of the LF
estimate under the conditions imposed by the use of photometric
redshifts instead of spectroscopic distances. We generate mock
MUNICS-like catalogs following a non-evolving LF and compare the
output of the LF estimate as described above with the input LF. Note
that we can only test the effects of random errors in the distances
and not the influence of systematically wrong distances as those shown
in Fig.~\ref{f:zold-znew}.

The simulated LF follows the form of th local 2MASS $K$-band LF of
\citet{Kochaneketal01}. We assume that there is no redshift evolution
in brightness or number density. The extrapolation of the flux to the
rest-frame $K$-band was done using the mean $k$-correction of the
models used for determining photometric redshifts
(Fig.~\ref{f:model-seds}). 

To investigate the influence that errors in the photometric redshifts
have on recovering the form of the LF, we simulate four data sets with
errors $\Delta z$ drawn from a Gaussian of width $0.02$, $0.055$ (the
value found from the comparison with spectroscopic redshifts in
Fig.~\ref{f:zz}), $0.1$, and $0.2$. The simulations are repeated 10
times, and the LF is extracted in four redshift bins, $0.4<z<0.6,
0.6<z<0.8, 0.8<z<1.0,$ and $1.0<z<1.2$. The results are shown in
Fig.~\ref{f:lfk-fake}.

The overall impression from Fig.~\ref{f:lfk-fake} is that the form is
well reproduced, and that, as expected, the influence of the redshift
uncertainty is larger at small redshifts where the relative influence
on the distance is largest. The form of the LF only becomes
significantly biased when the redshift errors exceed $\Delta z \ga
0.1$, becoming a significant fraction of the redshift bin size. There,
the well-known effect of objects being preferably scattered away from
$L^*$, producing too many bright objects and too few lower luminosity
objects becomes apparent.  For the value $\Delta z = 0.055$ (upper
right panel in Fig.~\ref{f:lfk-fake}), the form is only slightly
affected in the lowest redshift bin, $0.4<z<0.6$.

From the last said we conclude that the shape of the luminosity
function is not significantly biased as long as the photometric
redshifts scatter symmetrically around the true redshifts \citep[see
also][]{SCSK96} and their errors are (significantly) smaller than the
bin size in $z$ over which the luminosity function is averaged.

\subsection{Influence of the incompleteness corrections}

The corrections for incompleteness involve the volume correction,
$1/\Vmax$, and the correction for the detection incompleteness at
fainter magnitudes, $1/P(m_{K,i})$.

Given our magnitude limits, both corrections do not play any role at
magnitudes brighter than $M_K^* \approx -23.5$. At fainter magnitudes
both corrections contribute significantly. The \Vmax\ term becomes
dominant at the faintest levels (as we exclude objects with
$P(m_{K,i}) < 0.75$).

In the lowest redshift bin the \Vmax\ term is ill-determined resulting
in a steepening in the LF's very faint end. The reason is the redshift
error associated with photometric redshifts and that the approximation
of the visibility function as that of a point source breaks down at
low redshifts where the objects are well-resolved. The steepening at
of the LF can be explained in terms of low luminosity galaxies having
true redshifts just lower than the bin's lower bound, which are
scattered upwards in $z$ by the photometric redshift estimate and just
make it into the bin. As those galaxies are close to the detection
limit, the \Vmax\ test predicts them to be detectable only in a very
small volume of the bin and therefore the \Vmax\ correction becomes
large (factors of 5 and above), producing the observed behavior.
Therefore our data do not allow us to place constraints on the faint
end slope of the LF, even in our lowest redshift bin.


\section{Results}
\label{sec:lf-results}

In this section we present the results on the evolution of the
rest-frame $K$-band luminosity function to $z \leq 1.2$.

In Fig.~\ref{f:lfk-final} we plot the rest-frame $K$-band luminosity
function obtained from the present MUNICS dataset. The plot shows the
uncorrected values as well as the \Vmax\ and incompleteness corrected
ones. For comparison, we also show the local $z=0$ 2MASS $K$-band LF
\citep{Kochaneketal01} and a Schechter function evolved with $z$ using
the best-fit values for the evolution parameters $\mu = 1/\Phi^*(0)
d\Phi^*/dz$ and $\nu = dM^*/dz$ from
Sect.~\ref{sec:likel-analys-lumin}. Note that these are not
independent Schechter fits in each redshift bin but rather a global
estimate of the change in $\Phi^*$ and $L^*$ with redshift (see
below).

Taken at face value, it is apparent that the total rest-frame $K$-band
LF evolves only mildly to $z = 1.2$. This is in qualitative agreement
with the finding of \citet{CSHC96} from a much smaller spectroscopic
sample, where the authors found that the $K$-band LF in the redshift
bin $0.6 < z < 1$ was consistent with the one at $0 < z < 0.2$ (using
an Einstein-de-Sitter cosmology).

This is further demonstrated by the total $K$-band luminosity density
shown in Fig.~\ref{f:lfk-lumdens}. Also here there is only mild
evolution with redshift within our sample, with the luminosity density
scaling as $d\log\rho_L / dz = 0.24$.

Note that to compute the previous quantity, we have to assume
something about the luminosity in objects at the faint end of the
luminosity function which are beyond our magnitude limits. Since we
have no handle on the slope of the faint end from our data, we assume
that the local slope of $\alpha =-1.09$ holds. We use the fraction of
the total luminosity sampled by the data to the completeness limit
$M_{\mathrm{min}}$,
\begin{equation}
  C_{\Phi} \; = \; \frac{ \int_{-\infty}^{M_{\mathrm{min}}}\Phi(M)dM }%
  {\int_{-\infty}^{+\infty}\Phi(M)\,dM },
\end{equation}
to correct for this effect in Fig.~\ref{f:lfk-lumdens}. This
correction is found to be smaller than a factor of 2 at all redshifts.

Again, we see clearly see the signature of cosmic variance in the
luminosity density measurements in the single Mosaic Fields compared
with the sample average. The values in the individual fields scatter
by $\sim 0.1$ to $\sim 0.15$ in $\log \rho_L$.

We can compare the luminosity density values obtained by summing over
the data with what is expected if the LF is well-represented by a
Schechter function, in which case the luminosity density is simply
given by $\rho_L = \Phi^* L^* \Gamma(2+\alpha)$. Using the best values
for the evolution parameters $\mu$ and $\nu$ from
Sect.~\ref{sec:likel-analys-lumin}, these expected values are shown in
Fig.~\ref{f:lfk-lumdens} as a solid line.

Our result on the evolution of the $K$-band LF is in marked contrast
to the results obtained for the $B$-band LF in the same redshift range
from the CFRS and CNOC surveys \citep{CFRS695,CNOC299} which
consistently find an increase in the $B$-band luminosity density
driven mainly by late spectral types.

At least some of this increase is due to fainter late-type galaxies at
(observer's frame) $B \ga 22$~mag which first appeared in faint blue
number counts \citep[see review by][]{Ellis97}. For example,
\citet{CSHC96} found that at $B \approx 24$, the population of
galaxies is a mixture of normal galaxies at modest redshifts and a
population of galaxies with a wide range of masses undergoing rapid
star formation which are spread out in redshift from $z = 0.2$ to at
least $z = 1.7$. The remaining part of the increase in blue luminosity
is due to luminosity evolution of normal field spirals and early-type
systems.

On the other hand, luminosity evolution is inevitable if the $K$-band
light traces the underlying old stellar population of galaxies, since
these stars obviously become younger with look-back time.  For the $K$
band, pure luminosity evolution predicts brightening by $\sim 0.6$~mag
as the average mass-to-light ratio evolves away from its local value
(see MUNICS~3; \citealp{PBZ96}).

In MUNICS~III we have shown a luminosity function based on the VRIJK
photometry available at the time. Although we did not present a
quantitative analysis of the luminosity function evolution in that
paper, it appeared that the LF is consistent with the local shape over
the entire range in $z$, while we do see departure from the local
shape in the present analysis. The explanation is found to be a weak
excess of objects at high $z$ in the present BVRIJK data (see
Fig.~\ref{f:zold-znew}) discussed above, which is enough to explain the
brightening seen at the high-luminosity end. A comparison of the
luminosity density in the limits adopted in MUNICS~III (Fig.~3) indeed
yields only small differences, of the order of 5 to 10 per cent.  We
wish to emphasize that a non-evolving LF does not imply a non-
evolving galaxy population (at least as long as we can only measure
the bright ($L>L^*$) part of the LF). Indeed, MUNICS~III argued that
density evolution is counter balancing luminosity evolution, leading
to a constant overall LF. In the current analysis based on broader
wavelength coverage and better spectroscopic calibration, the balance
between luminosity and density evolution is altered to slightly more
luminosity evolution compared to MUNICS~III, but the principal results
remain the same as will be shown in a detailed quantitative analysis
below.


\section{Likelihood analysis of luminosity function evolution}
\label{sec:likel-analys-lumin}

We define the amount of number density and luminosity evolution in our
sample as the change of $\Phi^*$ and $M^*$ with redshift,
respectively, in the Schechter parametrization of the LF. To quantify
this, we apply two different estimators for the evolution of the LF
with redshift, one operating on the data binned in magnitude and
redshift, and one operating solely on individual objects, i.e.\ 
unbinned individual magnitude and redshift $\{M_i, z_i\}$ pairs.

In both cases we use the \citet{Schechter76} parametrization of the
luminosity function,

\begin{equation}
  \Phi(L)\,dL \; = \; \frac{\Phi^*}{L^*} \, \left( \frac{L}{L^*}
  \right)^\alpha \, \exp \left( - \frac{L}{L^*} \right) \, dL
  \label{eq:schechter}
\end{equation}

where $L^*$ is the characteristic luminosity, $\alpha$ the faint-end
slope, and $\Phi^*$ the normalization of the luminosity function. The
corresponding equation in absolute magnitudes reads

\begin{eqnarray}
  \nonumber
  \Phi(M)\,dM & = & \frac{2}{5} \, \Phi^* \, \ln 10 \: \times \: 
  10^{0.4(M^*-M)(1+\alpha)} \\
  \nonumber
  & \times & \exp \left( - 10^{0.4 (M^*-M)} \right)\,dM .
  \label{eq:schechter2}
\end{eqnarray}

To estimate the rate of evolution in this parametrization with
redshift, we define evolution parameters $\mu$ and $\nu$ as follows:
\begin{eqnarray}
  \nonumber
  \Phi^* \, (z) & = & \Phi^* \, (0) \: \left( 1 \, + \, \mu z \right)
  , \\
  M^*    \, (z) & = & M^*    \, (0) \, + \, \nu z , \; \mathrm{and} 
\label{eq:evol}\\
  \nonumber
  \alpha \, (z) & = & \alpha \, (0) \;\; \equiv \;\; \alpha .
\end{eqnarray}

And therefore,
\begin{eqnarray}
  \nonumber
  \mu & = & \frac{1}{\Phi^*(0)} \frac{d\Phi^*(z)}{dz} \\
  \nu & = & \frac{dM^*(z)}{dz} .
\end{eqnarray}

Note that the faint end slope of the luminosity function cannot be
determined from our data, thus we leave the faint-end slope $\alpha$
of the Schechter luminosity function fixed.

For the values of $M^*(0)$ and $\Phi^*(0)$ we use the values given by
\citet{Kochaneketal01} for 2MASS,
\begin{eqnarray}
\nonumber
\Phi^*(0) & = & 1.16 \times 10^{-2} \,\,\, h^3 \, \mathrm{Mpc^{-3}} \\
\nonumber M^*(0)   & = & -23.39 + 5\,\,\log\,h \\
\nonumber \alpha    & = & -1.09 .
\end{eqnarray}

\subsection{A $\chi^2$-based approach using binned data}

First, we apply a $\chi^2$-based estimator of the redshift evolution
of the near-infrared luminosity function. The same estimator is
applied in MUNICS~V to a smaller but purely spectroscopic sample.

To quantify the redshift evolution of $\Phi^*$ and $M^*$ we compare
our luminosity function data in \textit{all} redshift bins with the
local Schechter function evolved according to equation (\ref{eq:evol})
to the appropriate redshift. We do this for a grid of values of $\mu$
and $\nu$, and calculate the value of $\chi^2$ for each grid point
according to

\begin{equation}
  \chi^2 \, (\mu, \nu) \; = \; \frac{1}{n} \, \sum_{i=1}^N \, \frac{\left[
  \phi (M_i, z_i) - \Phi_{\mu \nu} (M_i, z_i ) \right]^2}{\sigma_i^2} ,
\end{equation}

where the sum is taken over all bins $i$, $\phi (M_i, z_i)$ is the
measured value of the luminosity function at median redshift $z_i$ in
the magnitude bin centered on $M_i$, and $\Phi_{\mu \nu} (M, z)$ is
the value of the local Schechter function at magnitude $M$ evolved
according to the evolution model defined in equation (\ref{eq:evol})
to the redshift $z$. Furthermore, $\sigma_i$ is the RMS error of the
measured luminosity function value, and $n$ is the number of free
parameters, i.e.\ the number of data points used minus the number of
parameters derived from the fitting, 23 in our case.

We use the data binned as presented in Fig.~\ref{f:lfk-final},
spanning $0.4<z<1.2$. However, we exclude all data points with a total
correction factor (incompleteness and $V/\Vmax$ correction) larger
than two. The errors of the measured LF, $\sigma_i$, are calculated
from the number of objects in the bin and assumed to follow Poisson
statistics.

The results of this approach are shown in the left panel of
Fig.~\ref{f:lfk-lik} and the most likely values of $\mu$ and $\nu$ are
listed in Table~\ref{t:lf-results}, along with the results from the
analysis of the spectroscopic sample in MUNICS~V.

\subsection{A likelihood-based approach using unbinned data}

The luminosity function is very steep at its bright end. When the
number of very bright (and therefore rare) galaxies in the sample is
small, binning of these data in magnitude must be taken with
skepticism.  As the distribution of objects within any bin will not be
uniform and we are dealing with small numbers of objects, the choice
of bin centers and widths might influence the outcome of the analysis.
Hence we develop a maximum likelihood based method avoiding having to
bin the data to determine the values of the evolution parameters $\mu$
and $\nu$.

Maximum likelihood methods (using binned and unbinned data) have been
used in the past by many authors to construct luminosity functions and
derive their parameters in various parametric and non-parametric ways
\citep[see, e.g.,][and references therein]{STY79,EEP88,LCRS96a}. These
methods usually suffer from the inconvenience that the normalization
has to be determined independently from the shape of the LF. The
method presented here allows the shape and normalization (evolution)
parameters to be determined simultaneously.

We start with a set of \Nobs\ observed galaxies of absolute magnitudes
$M_i$ and redshifts $z_i$. Assuming a luminosity function, $\Phi_{\mu
  \nu}(M,z)$, and parameterizing its redshift evolution as described
before (Eqs.~\ref{eq:schechter} and \ref{eq:evol}), one can write the
probability of observing galaxy $i$ having the properties $M_i$ and
$z_i$ as

\begin{equation}
  P(M_i,z_i|\Phi_{\mu \nu})\,dM\,dz \; = \; \frac{\Phi_{\mu \nu}(M_i, z_i)
                        (dV(z_i)/dz) }{\Nmodel}\,dM\,dz.
  \label{eq:like1}
\end{equation}

The normalization, \Nmodel, is the double integral over magnitude and
redshift of the LF, and hence equal the total number of galaxies
expected in the sample if the LF evolves according to our
parametrization:
\begin{equation}
  N_{\mathrm{model}} \; = \; \int_{\zmin}^{\zmax} \, \int_{\Mmin(z)}^{\Mmax}
                          \Phi_{\mu \nu}(M,z) \, \frac{dV}{dz}
                          \,dM\,dz
  \label{eq:norm}
\end{equation}

Here we need to take the redshift and magnitude limits of the survey
into account. The galaxies are taken to be in the redshift range
$\zmin\ < z < \zmax$, and to have absolute magnitudes $\Mmax < M <
\Mmin(z)$.

The lower magnitude limit is clearly a function of $z$, and can be
written as
\begin{equation}
  \Mmin(z) \; = \; m_{\mathrm{min}} - 2.5\,\log(1+z) - k_c(z) - D_m(z) ,
\end{equation}
where $m_{\mathrm{min}}$ is the completeness limit of the survey in
apparent magnitude, $k_c(z)$ is the (type-dependent) $k$-correction at
redshift $z$, and $D_m(z)$ is the distance modulus. The
$k$-corrections for each object are obtained from its best-fitting SED
(see Sect.~\ref{sec:phot-redsh}).

Since we do not want to deal with incompleteness corrections here, we
set
\begin{eqnarray}
  \nonumber
  m_{\mathrm{min}} & = & 18.7 \quad \mathrm{in~the}~K~\mathrm{band}\\
  \nonumber
  \zmin & = & 0.4\\
  \nonumber
  \zmax & = & 1.0 
\end{eqnarray}
where MUNICS is to be considered complete (see MUNICS~IV for the
detailed completeness analysis).

Note that Eq.~\ref{eq:like1} cannot be used to determine the
normalization, $\Phi^*$, or its evolution with redshift, $\mu$, since
the dependence on $\Phi^*$ cancels out (see also the method presented
in \citealp{LCRS96a}). A handle on the normalization can be obtained
by observing that the total number of objects in the sample, \Nobs,
must be reproduced by any successful model, $\Phi_{\mu \nu}$, leading
to a boundary condition for Eq.~\ref{eq:like1}. The total number of
objects in the sample is by itself subject to uncertainty, obeying at
least Poisson statistics (assuming that the total survey volume is
large enough such that fluctuations due to large scale structure are
averaged out). Instead of a boundary condition, we thus add a second
term to Eq.~\ref{eq:like1} taking the Poisson noise in \Nobs\ into
account to obtain the final likelihood function
\begin{equation}
  \ln \Lik \; = \;\mathop{\sum}_{i=1}^{\Nobs} \ln P(M_i,z_i|\Phi_{\mu \nu})
                - \frac{1}{2}
                \left(\frac{\Nmodel - \Nobs}{\sqrt{\Nobs}}\right)^2 ,
\label{eq:likelihood}
\end{equation}

The results of this maximum likelihood approach are shown in the right
panel of Fig.~\ref{f:lfk-lik} and the most likely values of $\mu$ and
$\nu$ are listed in Table~\ref{t:lf-results}. The dashed line in
Fig.~\ref{f:lfk-final} shows a Schechter function evolved with $z$
using these numbers.


\section{Discussion}\label{sec:discussion}

The $\chi^2$ estimates for $\mu$ and $\nu$ yield a mild decrease in
number density of 30\% to $z = 1$ ($\mu=-0.30 \pm 0.21$) accompanied
by a brightening of the galaxy population by $0.64 \pm 0.21$~mag. The
uncertainties of the parameters are large, though, because of the
relatively large uncertainties of the values of the LF in the
individual bins, and because the Schechter LF parameters are almost
degenerate if the faint end below $L^*$ cannot be measured (at the
steep bright end, negative density evolution can be counterbalanced by
brightening).

The results from the likelihood estimator are fully consistent with
the $\chi^2$ estimates, with their uncertainties being much smaller,
though (see comparison in Table~\ref{t:lf-results}). The likelihood
estimator gives 25\% number density evolution ($\mu = -0.25 \pm 0.05$)
and a brightening of $0.53 \pm 0.07$~mag. $\mu = 0$ is excluded at
high confidence ($\sim 4\sigma$). The reason for the much steeper
contours of the likelihood surface compared to the $\chi^2$ estimator
come from the tightness of the constraint in total number of galaxies
(Eq.~\ref{eq:norm}). This number is very sensitive to small changes in
the Schechter parameters, and therefore provides a very strong
constraint on the allowed region in parameter space.

It well worth noting that the parameter estimates obtained by both
methods presented here agree with the best estimate obtained in
MUNICS~V from the spectroscopic sample alone. This makes us confident
that the results presented here based on photometric redshifts are,
indeed, robust. The formal uncertainties we derive in this work are,
however, much smaller than those of the analysis of the purely
spectroscopic sample. This is entirely due to the sample size which is
a factor of 10 larger than the spectroscopic one in MUNICS~V.  The
results presented here are also consistent with the recent estimate
based on a spectroscopic sample by \citet{K20-03}, finding brightening
by $0.54 \pm 0.12$ magnitudes to $z=1$ and at most a mild decrease ($<
30\%$) in number density of red luminous objects.

Obviously, the results of both estimators crucially depend on the
parameters assumed for the local ($z=0$) values of the luminosity
function. We estimate the impact on the parameters due to this by
using a direct fit of a Schechter function to the lowest redshift bin
($0.4<z<0.6$) of the MUNICS sample itself (upper left panel in
Fig.~\ref{f:lfk-final}) as a reference instead of the $z=0$
measurement. The results are effectively unchanged, being consistent
within the $1\sigma$ uncertainties in both cases, in fact being almost
identical for the $\chi^2$ estimator.

Another possible source of systematic error is possible incompleteness
in the MUNICS photometric catalogs at high redshift. This would effect
the validity of the spectroscopic results in MUNICS~V alike since the
objects for spectroscopy are selected from the same photometric
catalog. The completeness analysis presented in MUNICS~IV concluded
that detection probabilities are high independent of the sources'
radial profile out to $z \sim 1$, but that photometry suffers from
severe biases for bulge-dominated systems where the lost light
fraction can be high for the highest luminosity objects.  For lower
luminosity objects around $L^*$ the photometry recovers the total
magnitudes well. This is a consequence of the relation between surface
brightness and total luminosity for elliptical galaxies. This effect
scatters very bright objects towards lower luminosities. Since bright
objects are very rare, the effect on the shape of the LF around $L^*$
is small, but can be severe at $L \gg L^*$. The total magnitude of
disk dominated systems are well recovered regardless of total
luminosity.  Unfortunately, one cannot correct for this effect without
assuming a distribution of morphological types or suitable high
resolution imaging which is still unavailable for a sample of this
size.

To investigate whether our results are sensitive to these effects, we
repeat the analysis described above using a number of restricted
redshift ranges. Firstly, lowering the upper limit in redshift down to
$z=0.8$, we find that the estimate for the density evolution changes
by not more than $\Delta\mu=0.08$ with the evolution in luminosity
being practically unaffected. The error in $\mu$ increases by $\sim
50\%$, though, because of the smaller number of objects in the
restricted-redshift samples. Secondly, we exclude objects at $z<0.6$,
since the form of the LF in this bin might be altered at the bright
end by the use of photometric redshifts (see
Sect.~\ref{sec:monte-carlo-simul} and Fig.~\ref{f:lfk-fake}). The
result of our analysis did not change significantly, since we use all
objects to estimate the value of the evolution parameters rather than
relying on determinations in individual redshift bins.

Finally, the faint end slope, $\alpha$, plays a role in our analysis.
The current sample does not go deep enough to constrain $\alpha$,
which is therefore kept fixed at its local value. The $\chi^2$
estimator is not very sensitive to the actual value of $\alpha$, since
the faint part of the LF is not sampled much, and the bins are
weighted by their $S/N$, such that the total $\chi^2$ is dominated by
the $L^*$ region (where objects are already numerous and \Vmax\ 
corrections are not yet important).

The maximum likelihood estimator is, however, sensitive to the
adapted value of $\alpha$. Again, this is because of the tightness of
the total-number constraint and the fact that faint objects dominate
in number (for all likely values of $\alpha$). We can estimate the
size of the systematic error associated with this by using a value of
$\alpha = -1$ and repeating the analysis. The model predicts fewer
galaxies in this case, and therefore the amount of density evolution
required is lower, $\mu = 0.15$. On the other hand, the LF in our
lowest redshift (still $z = 0.5$) bin is inconsistent with $\alpha=1$
(fixing $\Phi^*$ to its local value) at the $\sim 2\sigma$ level.


\section{Summary}\label{sec:summary}

We present a measurement of the evolution of the rest-frame $K$-band
luminosity function to $z \sim 1$ using a sample of $K$-selected
galaxies drawn from the MUNICS dataset. 

Distances and absolute $K$-band magnitudes are derived using
photometric redshifts from template SED fits to BVRIJK photometry.
These are calibrated using $> 500$ spectroscopic redshifts from a
spectroscopic follow-up program described in MUNICS~V. A method for
obtaining templated SEDs matching the properties of the sample is
presented and the photometric redshift estimation is explained. We
obtain redshift estimates having a rms scatter of $0.055$ and no mean
bias.

\begin{figure}
  \epsscale{0.45}
  \plotone{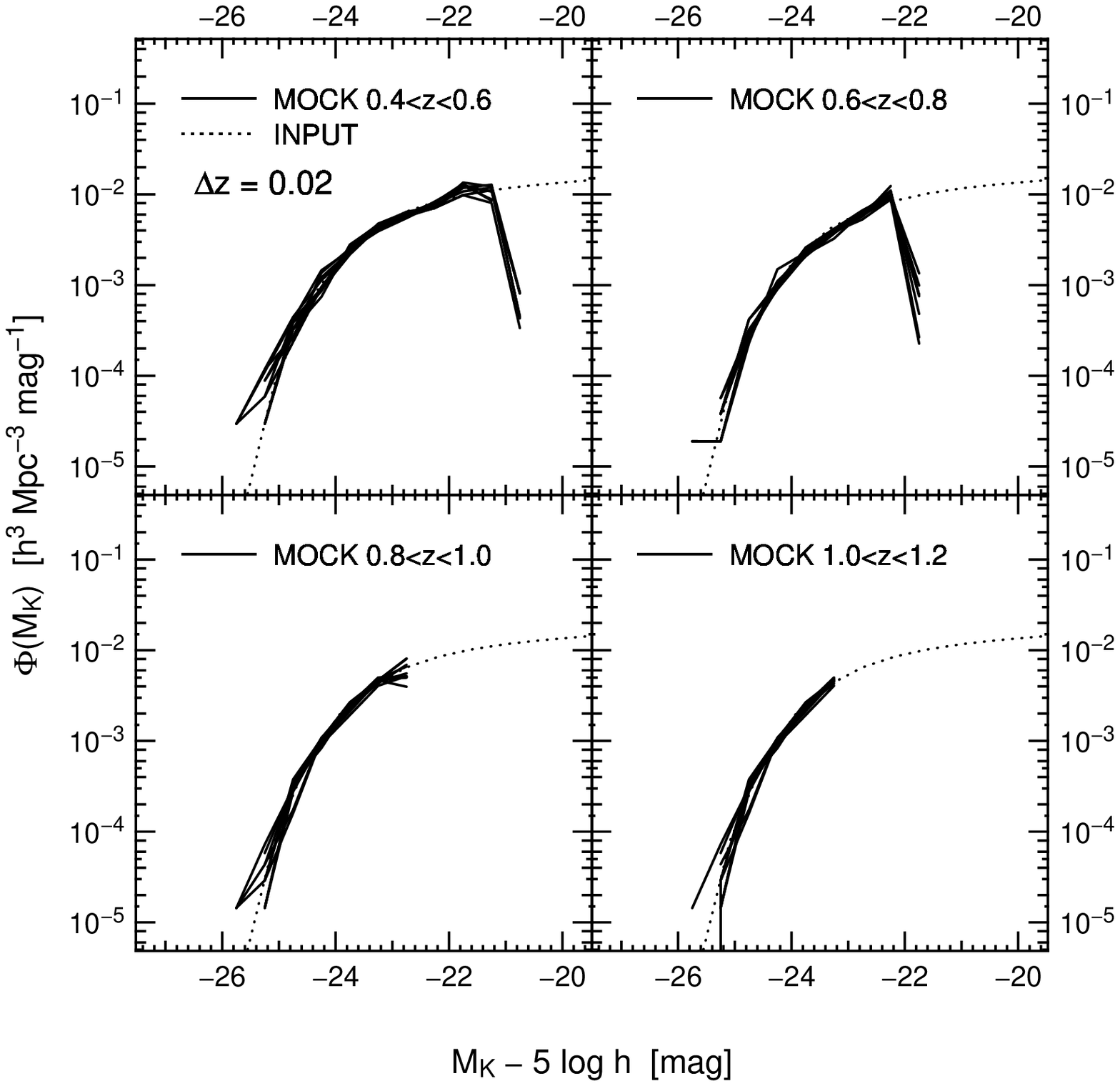}
  \plotone{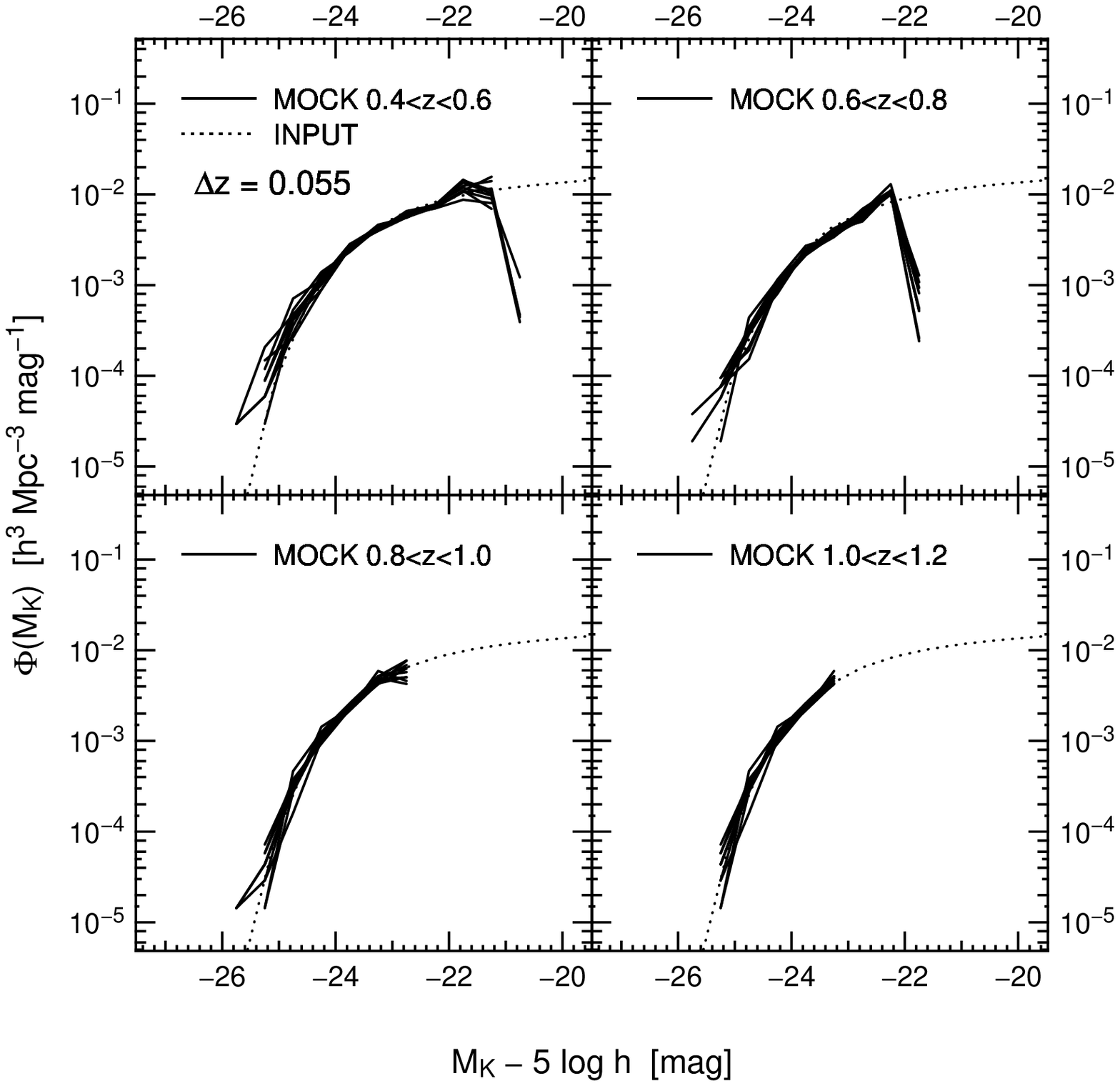}
  \plotone{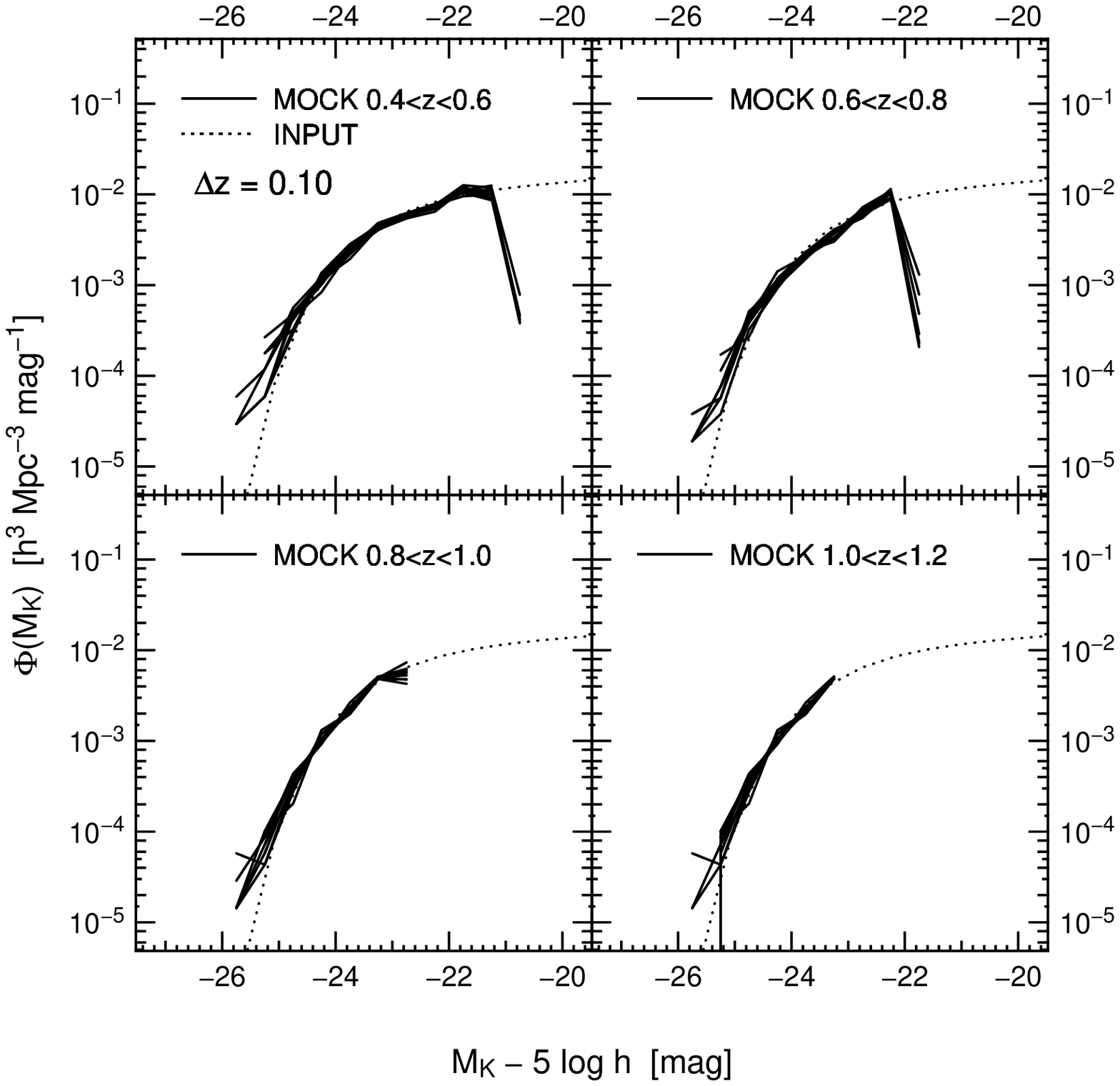}
  \plotone{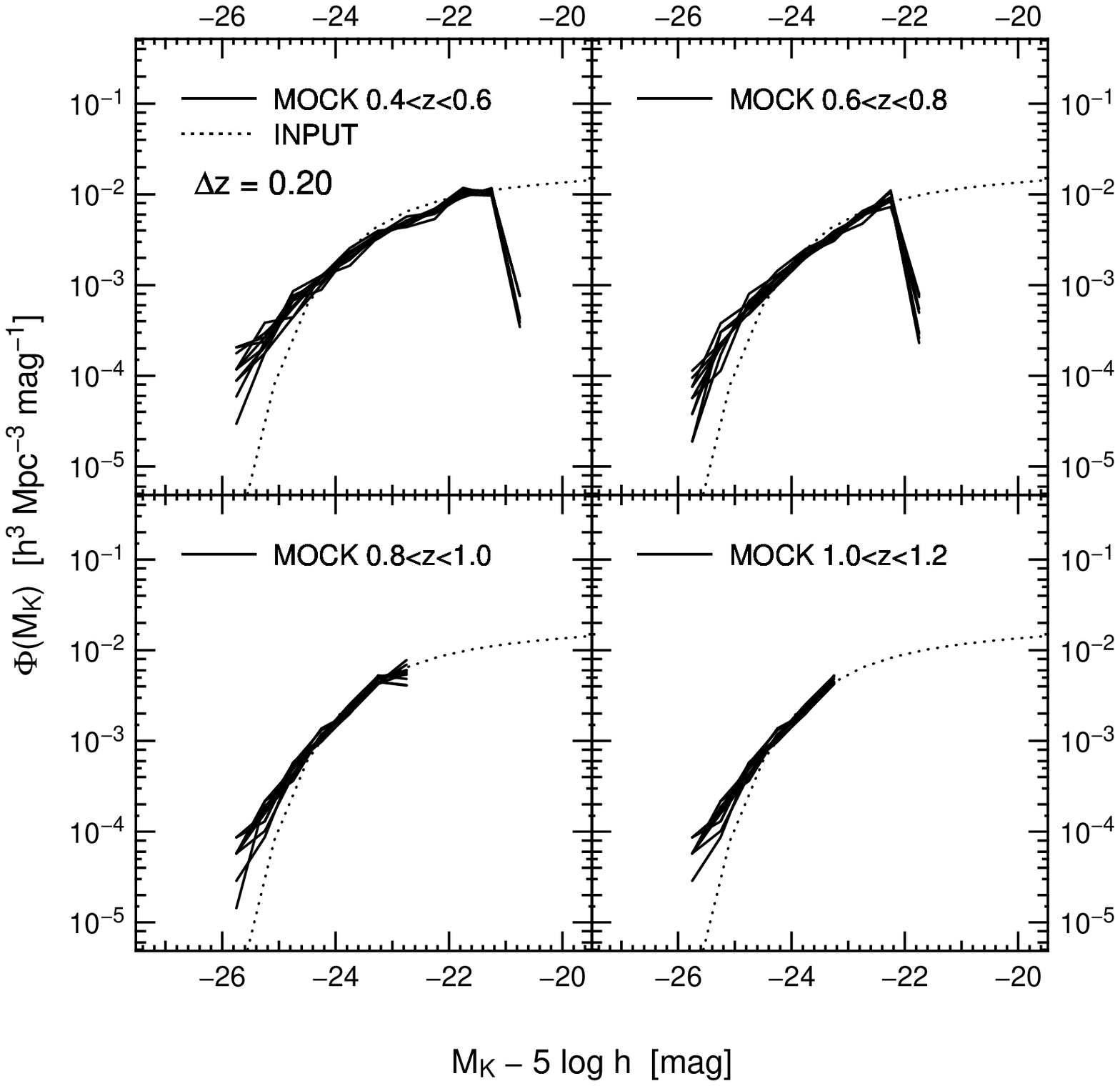}
  \caption{Monte-Carlo realizations of the luminosity function to test
    the susceptibility to random errors in the redshift determination
    due to the use photometric redshifts. The LFs were simulated with
    redshift errors $\Delta z$ drawn from a Gaussian of width $0.02$
    (upper left), $0.055$ (upper right; this corresponds to the rms
    redshift error in our sample), $0.1$ (lower left), and $0.2$
    (lower right). In each panel, the LF is shown in four redshift
    bins, $0.4<z<0.6, 0.6<z<0.8, 0.8<z<1.0,$ and $1.0<z<1.2$ (see
    text).  Each simulation was repeated 10 times, assuming a
    non-evolving luminosity function of the form the local 2MASS
    $K$-band LF (dotted line). The simulated LF is generally recovered
    well, with the known effect that objects are being preferably
    scattered away from $L^*$ becoming apparent at high rms redshift
    errors.}
  \label{f:lfk-fake}
\end{figure}

We use Monte-Carlo simulations to investigate the influence of the
errors in distance associated with photometric redshifts on our
ability to reconstruct the shape of the LF (Fig.~\ref{f:lfk-fake}).

\begin{figure}
  \epsscale{0.95}
  \plotone{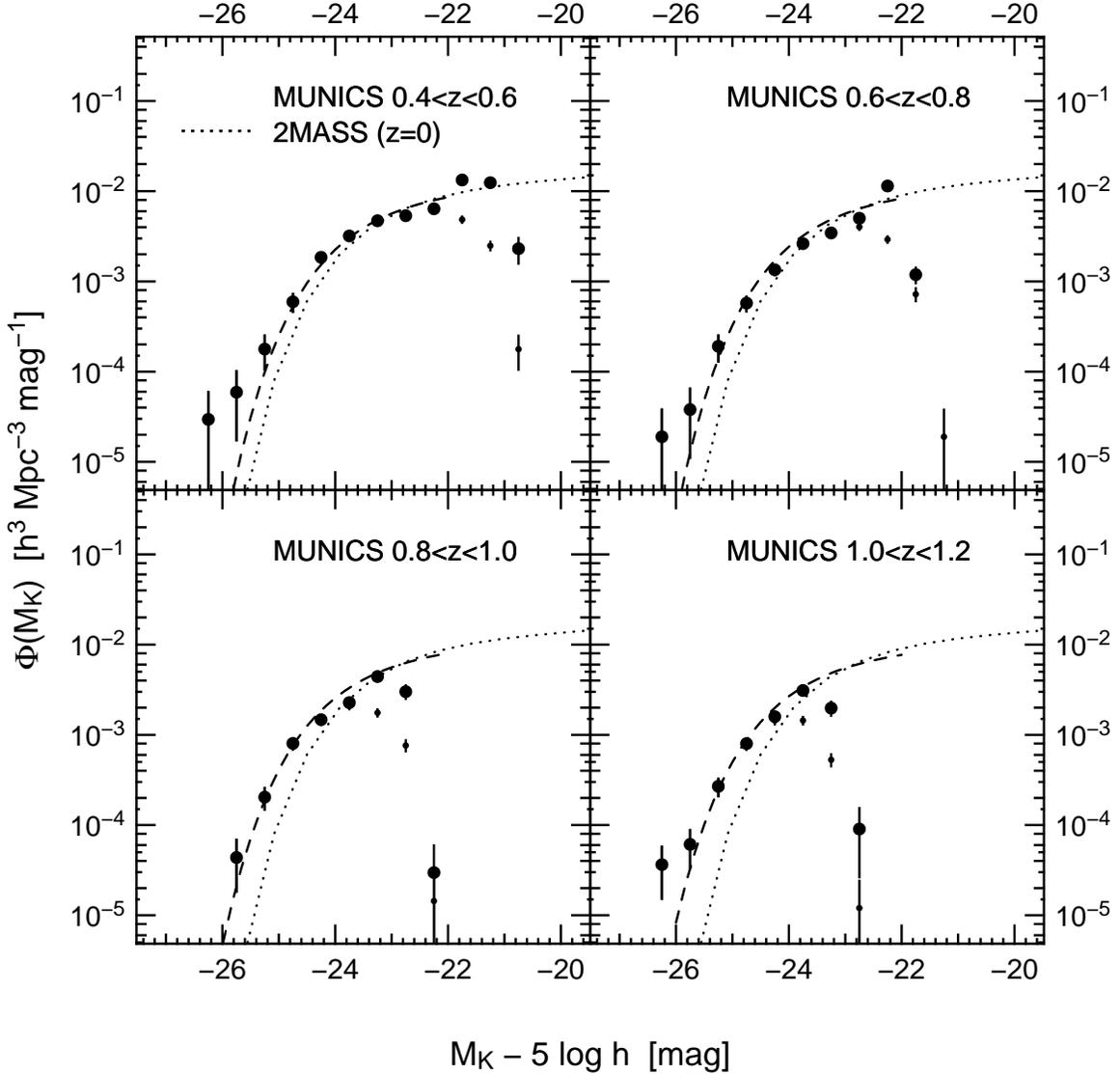}
  \caption{The final rest-frame $K$-band luminosity 
    function from the MUNICS sample in four redshift bins, $0.4<z<0.6,
    0.6<z<0.8, 0.8<z<1.0,$ and $1.0<z<1.2$. The small filled symbols
    denote the uncorrected data, the large filled symbols show the
    final corrected data. Error bars are Poisson errors on the number
    of objects in each ${M_K,z}$ bin. The dotted line denotes the
    local 2MASS $K$-band LF published by \citet{Kochaneketal01}.  The
    dashed line shows a Schechter function evolved with $z$ using the
    best-fit values for the evolution parameters $\mu =-0.25$ and $\nu
    = -0.53$ from Sect.~\ref{sec:likel-analys-lumin}. Note that these
    are not independent fits in each redshift bin but rather a global
    estimate of the change in $\Phi^*$ and $L^*$ with redshift (see
    text).}
  \label{f:lfk-final}
\end{figure}

\begin{figure}
  \epsscale{0.6}
  \plotone{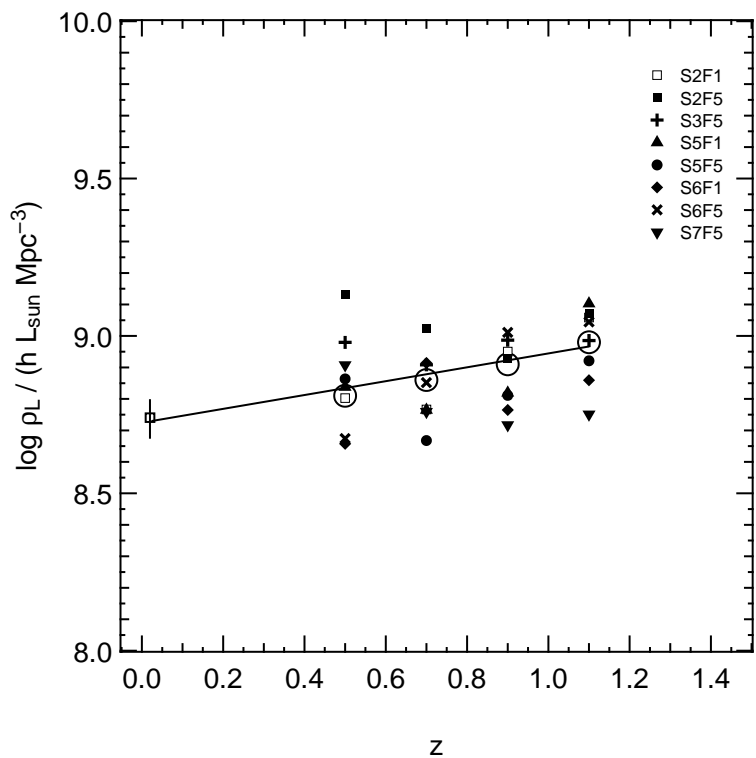}
  \caption{The rest-frame $K$-band luminosity density at $0.4 < z 1.2$.
    Different symbols denote the different survey patches (Mosaic
    Fields). The large open circles are average values. The point at
    $z=0.02$ is taken from \citet{Kochaneketal01}. The solid line
    represents the expected total luminosity density using the
    best-fit values for the evolution parameters $\mu = -0.25$ and
    $\nu = -0.53$ from Sect.~\ref{sec:likel-analys-lumin}.}
  \label{f:lfk-lumdens}
\end{figure}

\begin{figure}
  \epsscale{1}
  \plottwo{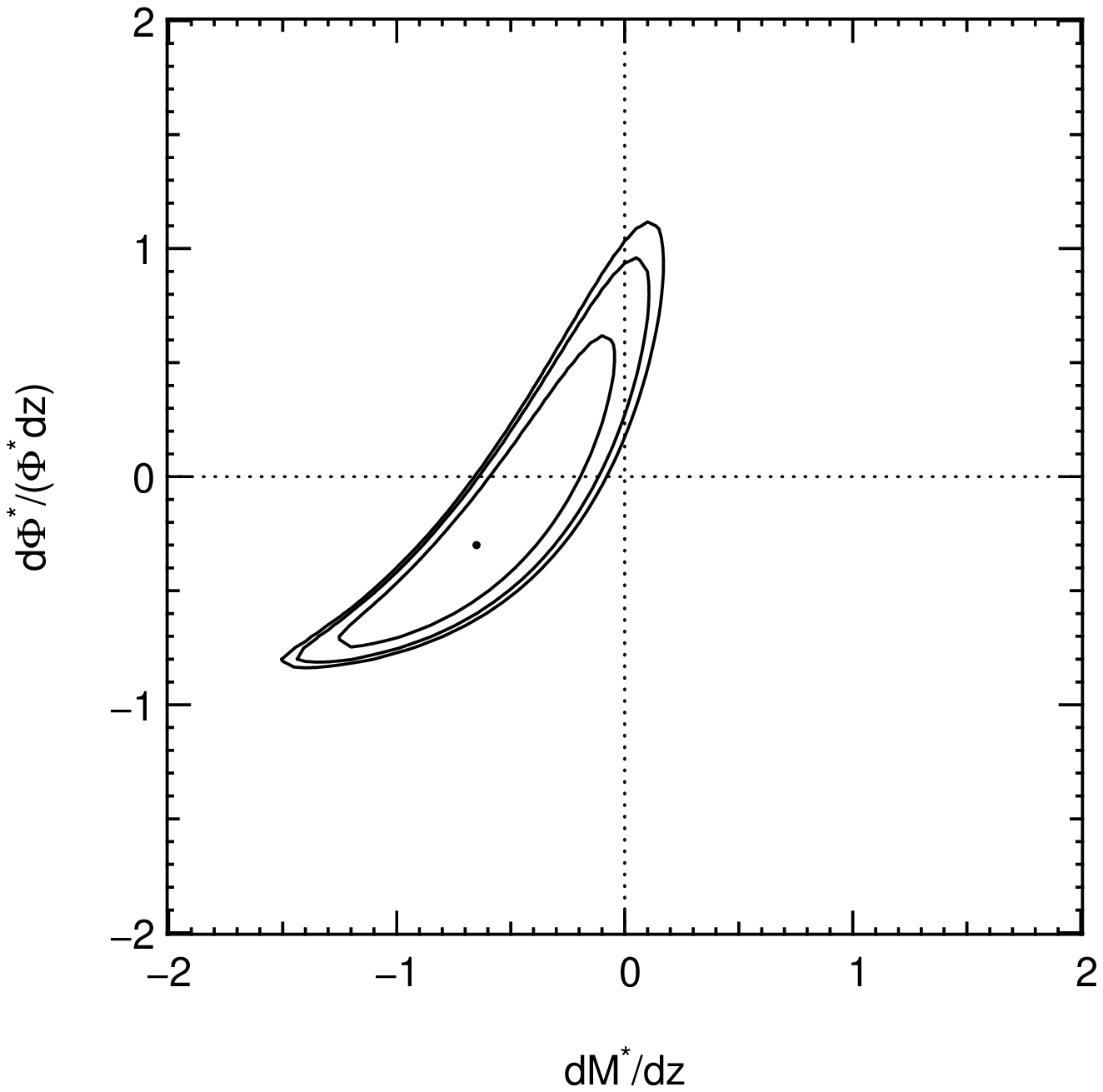}{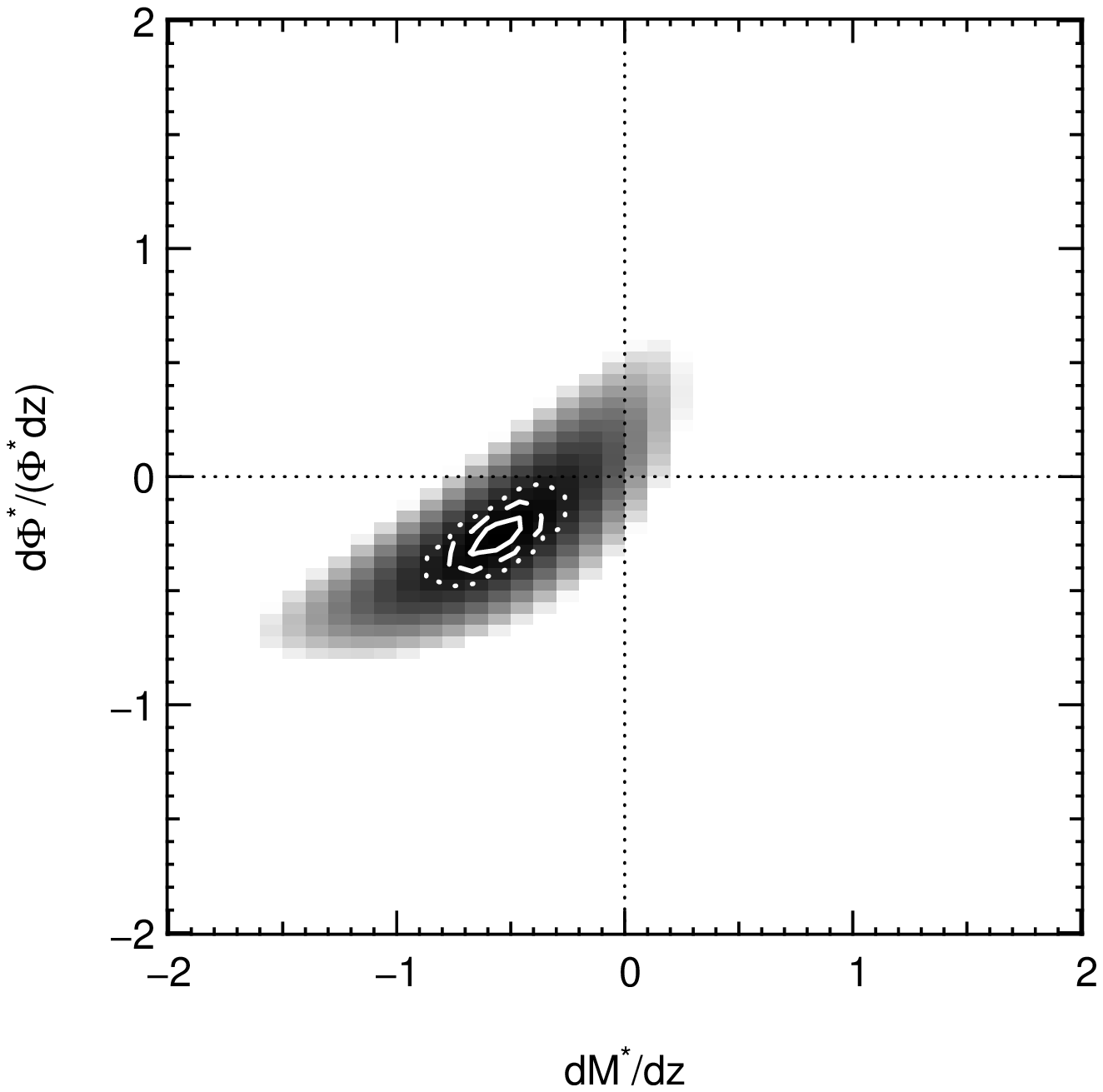}
  \caption{$\chi^2$ contours (left panel) and likelihood contours 
    (right panel) for the luminosity function evolution parameters
    (see text). The contours shown correspond to the $1\sigma$,
    $2\sigma$, and $3\sigma$ confidence levels.}
  \label{f:lfk-lik}
\end{figure}

We construct the rest-frame $K$-band LF in four redshift bins spanning
$0.4<z<1.2$ (Fig.~\ref{f:lfk-final}) and use two different estimators
to derive likely values for the evolution of $\Phi^*$ and $M^*$ with
redshift.  While the first estimator relies on the value of the
luminosity function binned in magnitude and redshift, the second
estimator uses the individually measured $\{M,z\}$ pairs alone. Our
results are shown in Fig.~\ref{f:lfk-lik} and are summarized in
Table~\ref{t:lf-results}. In both cases we obtain a mild decrease in
number density by $\sim 25\%$ to $z=1$ accompanied by brightening of
the galaxy population by $0.5$ to $0.7$~mag. These results are fully
consistent with an analogous analysis using only the spectroscopic
MUNICS sample (MUNICS~V).

Among the sources of systematic errors are the local values of the
parameters of the LF, especially the faint end slope, and possible
incompleteness and biases in the photometry (discussed in detail in
MUNICS~IV) present in the sample. We conclude that these have some
influence on our analysis, sometimes affecting the two estimators in
different ways, but that these effects do not influence the general
conclusion of this work, namely that there is a mild decrease in total
number density of $20\%$ to $30\%$ to $z \sim 1$ and brightening by
$0.5$ to $0.7$~mag, taking the systematic errors into account.

To interpret these results in terms of a picture of galaxy evolution
is not straight-forward. While the change in characteristic luminosity
is consistent with models of passive evolution of the stellar
populations, we also have to account for the evolution in number
density. Since the $K$-band light traces stellar mass, we may
attribute a change in $K$ band luminosity function either to a change
in the mass-to-light ratio alone (passive evolution) or to a
combination of a change in $M/L$ and in stellar mass. A change in
stellar mass can be due to star formation and/or due to merging and
accretion. These models cannot be discriminated on basis of the
luminosity function alone. Note, however, that in MUNICS~III we have
assumed a model of passive evolution of the mass-to-light ratio to
assign stellar masses to the galaxies in the sample. This model
assumed a maximum mass at the given $K$-band light by having all stars
form at $z=\infty$.  Analysis of the number density of objects as a
function of stellar mass showed that the total number of galaxies
changes only very little (like in this study), but that the number
density of systems more massive than $5\times 10^{10} h^{-2} \Msun$
does decline, and the number density of systems more massive than
$1\times 10^{11} h^{-2} \Msun$ declines even faster, a pattern which
is qualitatively consistent with hierarchical clustering, although
these models still tend to predict a more rapid decline in number
density of massive systems with redshift.

A PLE model for the evolution of $M/L$ for the whole galaxy population
is clearly too simplistic and provides only an upper envelope to the
stellar mass as a function of redshift. A more proper analysis has to
model the mass-to-light ratio of each object individually and take
star formation and dust extinction into account. We therefore postpone
further interpretation to a later paper (Drory et al.\ 2003, in
preparation; MUNICS~VI) where we use the photometry and spectroscopy
combined with stellar population models to investigate the evolution
of the stellar masses as a function of cosmic epoch.


\acknowledgments

This work was partly supported by the Deutsche Forschungsgemeinschaft,
grant SFB 375 ``Astroteilchenphysik''. ND acknowledges support by the
Alexander von Humboldt Foundation.  We would also like to thank the
Calar Alto staff for their long-standing support during many observing
runs over the last six years.  

This work is based on observations obtained at (1) the German-Spanish
Astronomical Center, Calar Alto, operated by the Max-Planck-Institut
f\"ur Astronomie, Heidelberg, jointly with the Spanish National
Commission for Astronomy, (2) McDonald Observatory, operated by the
University of Texas at Austin, and the Hobby - Eberly Telescope,
operated by McDonald Observatory on behalf of The University of Texas
at Austin, the Pennsylvania State University, Stanford University,
Ludwig-Maximilians-Universit\"at M\"unchen, and
Georg-August-Universit\"at G\"ottingen, and (3) the European Southern
Observatory, Chile, proposal number 66.A-0129 and 66.A-0123.



\begin{thebibliography}{}

\bibitem[\protect\citeauthoryear{{Bender}}{{Bender}}{2003}]{Bender03}
{Bender}, R. 2003, ApJ, submitted

\bibitem[\protect\citeauthoryear{{Bender} et~al.}{{Bender}
  et~al.}{2001}]{photred00}
{Bender}, R., et~al. 2001, in ESO/ECF/STScI Workshop on Deep Fields, ed.
  S.~Christiani (Berlin: Springer), 327

\bibitem[\protect\citeauthoryear{{Ben{\'\i}tez}}{{Ben{\'\i}tez}}{2000}]{Benite%
z00}
{Ben{\'\i}tez}, N. 2000, ApJ, 536, 571

\bibitem[\protect\citeauthoryear{{Bromley} et~al.}{{Bromley}
  et~al.}{1998}]{BPLK98}
{Bromley}, B.~C., {Press}, W.~H., {Lin}, H.,  \& {Kirshner}, R.~P. 1998, ApJ,
  505, 25

\bibitem[\protect\citeauthoryear{{Cimatti} et~al.}{{Cimatti}
  et~al.}{2002}]{K20-02}
{Cimatti}, A., et~al. 2002, A\&A, 381, L68

\bibitem[\protect\citeauthoryear{{Cowie} et~al.}{{Cowie} et~al.}{1996}]{CSHC96}
{Cowie}, L.~L., {Songaila}, A., {Hu}, E.~M.,  \& {Cohen}, J.~G. 1996, AJ, 112,
  839

\bibitem[\protect\citeauthoryear{{Drory} et~al.}{{Drory}
  et~al.}{2001a}]{MUNICS3}
{Drory}, N., {Bender}, R., {Snigula}, J., {Feulner}, G., {Hopp}, U.,
  {Maraston}, C., {Hill}, G.~J.,  \& {de Oliveira}, C.~M. 2001a, ApJ, 562, L111

\bibitem[\protect\citeauthoryear{{Drory} et~al.}{{Drory}
  et~al.}{2001b}]{MUNICS1}
{Drory}, N., {Feulner}, G., {Bender}, R., {Botzler}, C.~S., {Hopp}, U.,
  {Maraston}, C., {Mendes de Oliveira}, C.,  \& {Snigula}, J. 2001b, MNRAS,
  325, 550

\bibitem[\protect\citeauthoryear{{Efstathiou}, {Ellis}, \&
  {Peterson}}{{Efstathiou} et~al.}{1988}]{EEP88}
{Efstathiou}, G., {Ellis}, R.~S.,  \& {Peterson}, B.~A. 1988, MNRAS, 232, 431

\bibitem[\protect\citeauthoryear{{Ellis}}{{Ellis}}{1997}]{Ellis97}
{Ellis}, R.~S. 1997, ARA\&A, 35, 389

\bibitem[\protect\citeauthoryear{{Feulner} et~al.}{{Feulner}
  et~al.}{2003}]{MUNICS5}
{Feulner}, G., {Bender}, R., {Drory}, N., {Hopp}, U., {Snigula}, J.,  \&
  {Hill}, G.~J. 2003, MNRAS, 342, 506

\bibitem[\protect\citeauthoryear{{Folkes} et~al.}{{Folkes}
  et~al.}{1999}]{2dF99}
{Folkes}, S., et~al. 1999, MNRAS, 308, 459

\bibitem[\protect\citeauthoryear{{Gardner} et~al.}{{Gardner}
  et~al.}{1997}]{GSFC97}
{Gardner}, J.~P., {Sharples}, R.~M., {Frenk}, C.~S.,  \& {Carrasco}, B.~E.
  1997, ApJ, 480, L99

\bibitem[\protect\citeauthoryear{{Glazebrook} et~al.}{{Glazebrook}
  et~al.}{1995}]{GPMC95}
{Glazebrook}, K., {Peacock}, J.~A., {Miller}, L.,  \& {Collins}, C.~A. 1995,
  MNRAS, 275, 169

\bibitem[\protect\citeauthoryear{{Heidt} et~al.}{{Heidt} et~al.}{2003}]{FDF1}
{Heidt}, J., et~al. 2003, A\&A, 398, 49

\bibitem[\protect\citeauthoryear{{Heyl} et~al.}{{Heyl} et~al.}{1997}]{HCEB97}
{Heyl}, J., {Colless}, M., {Ellis}, R.~S.,  \& {Broadhurst}, T. 1997, MNRAS,
  285, 613

\bibitem[\protect\citeauthoryear{{Kinney} et~al.}{{Kinney}
  et~al.}{1996}]{KinneyCalzetti}
{Kinney}, A.~L., {Calzetti}, D., {Bohlin}, R.~C., {McQuade}, K.,
  {Storchi-Bergmann}, T.,  \& {Schmitt}, H.~R. 1996, ApJ, 467, 38

\bibitem[\protect\citeauthoryear{{Kochanek} et~al.}{{Kochanek}
  et~al.}{2001}]{Kochaneketal01}
{Kochanek}, C.~S., et~al. 2001, ApJ, 560, 566

\bibitem[\protect\citeauthoryear{{Lilly} et~al.}{{Lilly}
  et~al.}{1995a}]{CFRS95}
{Lilly}, S.~J., {Le F{\`e}vre}, O., {Crampton}, D., {Hammer}, F.,  \& {Tresse},
  L. 1995a, ApJ, 455, 50

\bibitem[\protect\citeauthoryear{{Lilly} et~al.}{{Lilly}
  et~al.}{1995b}]{CFRS695}
{Lilly}, S.~J., {Tresse}, L., {Hammer}, F., {Crampton}, D.,  \& {Le F{\`e}vre},
  O. 1995b, ApJ, 455, 108

\bibitem[\protect\citeauthoryear{{Lin} et~al.}{{Lin} et~al.}{1996}]{LCRS96a}
{Lin}, H., {Kirshner}, R.~P., {Shectman}, S.~A., {Landy}, S.~D., {Oemler}, A.,
  {Tucker}, D.~L.,  \& {Schechter}, P.~L. 1996, ApJ, 464, 60

\bibitem[\protect\citeauthoryear{{Lin} et~al.}{{Lin} et~al.}{1997}]{LYCE97}
{Lin}, H., {Yee}, H. K.~C., {Carlberg}, R.~G.,  \& {Ellingson}, E. 1997, ApJ,
  475, 494

\bibitem[\protect\citeauthoryear{{Lin} et~al.}{{Lin} et~al.}{1999}]{CNOC299}
{Lin}, H., {Yee}, H. K.~C., {Carlberg}, R.~G., {Morris}, S.~L., {Sawicki}, M.,
  {Patton}, D.~R., {Wirth}, G.,  \& {Shepherd}, C.~W. 1999, ApJ, 518, 533

\bibitem[\protect\citeauthoryear{{Liu} et~al.}{{Liu} et~al.}{1998}]{LGHO98}
{Liu}, C.~T., {Green}, R.~F., {Hall}, P.~B.,  \& {Osmer}, P.~S. 1998, AJ, 116,
  1082

\bibitem[\protect\citeauthoryear{{Madgwick} et~al.}{{Madgwick}
  et~al.}{2002}]{2dF02}
{Madgwick}, D.~S., et~al. 2002, MNRAS, 333, 133

\bibitem[\protect\citeauthoryear{{Maraston}}{{Maraston}}{1998}]{Maraston98}
{Maraston}, C. 1998, MNRAS, 300, 872

\bibitem[\protect\citeauthoryear{{Marzke}, {Huchra}, \& {Geller}}{{Marzke}
  et~al.}{1994}]{CFA94}
{Marzke}, R.~O., {Huchra}, J.~P.,  \& {Geller}, M.~J. 1994, ApJ, 428, 43

\bibitem[\protect\citeauthoryear{{Pozzetti}, {Bruzual A.}, \&
  {Zamorani}}{{Pozzetti} et~al.}{1996}]{PBZ96}
{Pozzetti}, L., {Bruzual A.}, G.,  \& {Zamorani}, G. 1996, MNRAS, 281, 953

\bibitem[\protect\citeauthoryear{{Pozzetti} et~al.}{{Pozzetti}
  et~al.}{2003}]{K20-03}
{Pozzetti}, L., et~al. 2003, A\&A, in press

\bibitem[\protect\citeauthoryear{{Ratcliffe} et~al.}{{Ratcliffe}
  et~al.}{1998}]{UKST98}
{Ratcliffe}, A., {Shanks}, T., {Parker}, Q.~A.,  \& {Fong}, R. 1998, MNRAS,
  293, 197

\bibitem[\protect\citeauthoryear{{Sandage}, {Tammann}, \& {Yahil}}{{Sandage}
  et~al.}{1979}]{STY79}
{Sandage}, A., {Tammann}, G.~A.,  \& {Yahil}, A. 1979, ApJ, 232, 352

\bibitem[\protect\citeauthoryear{{Schechter}}{{Schechter}}{1976}]{Schechter76}
{Schechter}, P. 1976, ApJ, 203, 297

\bibitem[\protect\citeauthoryear{{Schmidt}}{{Schmidt}}{1968}]{Schmi68}
{Schmidt}, M. 1968, ApJ, 151, 393

\bibitem[\protect\citeauthoryear{{Shapley} et~al.}{{Shapley}
  et~al.}{2001}]{SSADGP01}
{Shapley}, A.~E., {Steidel}, C.~C., {Adelberger}, K.~L., {Dickinson}, M.,
  {Giavalisco}, M.,  \& {Pettini}, M. 2001, ApJ, 562, 95

\bibitem[\protect\citeauthoryear{{Snigula} et~al.}{{Snigula}
  et~al.}{2002}]{MUNICS4}
{Snigula}, J., {Drory}, N., {Bender}, R., {Botzler}, C.~S., {Feulner}, G.,  \&
  {Hopp}, U. 2002, MNRAS, 336, 1329

\bibitem[\protect\citeauthoryear{{Subbarao} et~al.}{{Subbarao}
  et~al.}{1996}]{SCSK96}
{Subbarao}, M.~U., {Connolly}, A.~J., {Szalay}, A.~S.,  \& {Koo}, D.~C. 1996,
  AJ, 112, 929

\bibitem[\protect\citeauthoryear{{Szokoly} et~al.}{{Szokoly}
  et~al.}{1998}]{SSCM98}
{Szokoly}, G.~P., {Subbarao}, M.~U., {Connolly}, A.~J.,  \& {Mobasher}, B.
  1998, ApJ, 492, 452

\bibitem[\protect\citeauthoryear{{Takeuchi}, {Yoshikawa}, \&
  {Ishii}}{{Takeuchi} et~al.}{2000}]{TYI00}
{Takeuchi}, T.~T., {Yoshikawa}, K.,  \& {Ishii}, T.~T. 2000, ApJS, 129, 1

\end{thebibliography}

\newpage
\begin{table}
  \centering
  \begin{tabular}{llcll}
    Field & Area/arcmin$^2$ & $\qquad$ & Field & Area/arcmin$^2$\\
    \hline\hline
    S2F1 & 118.7 & & S5F5 & 107.6 \\
    S2F5 & 124.1 & & S6F1 & 130.8 \\
    S3F5 & 115.3 & & S6F5 & 140.2 \\
    S5F1 & 121.9 & & S7F5 & 139.1 \\
    \hline
  \end{tabular}
  \caption{List of the Mosaic Fields with best photometric 
    homogeneity, good seeing, and similar depth. The total area amounts
    to 997.7 square arcmin (0.28 square degrees). 
    There are $\sim 5000$ objects in the corresponding source catalog}
  \label{t:good-fields}
\end{table}

\begin{table}
  \centering
  \begin{tabular}{lll}
    Method & $\mu = \frac{1}{\Phi^*(0)}\frac{d \Phi^*}{d z}$ & 
             $\nu = \frac{d M^*}{d z}$ \\
    \hline\hline
    max.\ likelihood       & $-0.25 \pm 0.05\,(1\sigma)\;\;0.17\,(3\sigma)$
                           & $-0.53 \pm 0.07\,(1\sigma)\;\;0.18\,(3\sigma)$ \\
    $\chi^2$               & $-0.30 \pm 0.41\,(1\sigma)\;\;0.59\,(3\sigma)$  
                           & $-0.64 \pm 0.35\,(1\sigma)\;\;0.51\,(3\sigma)$ \\
    Spectroscopy (MUNICS~V)& $-0.35 \pm 0.5$ & $-0.7 \pm 0.3$ \\
    \hline
  \end{tabular}
  \caption{Max.\ likelihood and $\chi^2$ estimates (see text) for the evolution
  of the Schechter parameters of the rest-frame $K$-band luminosity function 
  to $z=1$. The results obtained by (\citealp{MUNICS5}; MUNICS~V) by 
  analyzing the spectroscopic sample alone are listed for comparison.}
  \label{t:lf-results}
\end{table}

\newpage

\end{document}